\documentclass[preprint,12pt]{elsarticle}

\usepackage[hidelinks]{hyperref}
\usepackage{graphicx}
\usepackage{subcaption}
\usepackage{amssymb}
\usepackage{amsmath}
\usepackage{mathrsfs}
\usepackage{multirow}
\usepackage[utf8]{inputenc}
\usepackage{ifthen}
\usepackage{cleveref}
\usepackage{listings}
\usepackage[ruled,vlined]{algorithm2e}
\usepackage{booktabs}
\usepackage{caption}
\usepackage{subcaption}
\usepackage{color}

\usepackage{xcolor}
\usepackage{xargs}
\usepackage[colorinlistoftodos,textsize=footnotesize]{todonotes}

\newcommandx{\unsure}[2][1=]{\todo[linecolor=red,backgroundcolor=red!25,bordercolor=red,#1]{#2}}
\newcommandx{\change}[2][1=]{\todo[linecolor=blue,backgroundcolor=blue!25,bordercolor=blue,#1]{#2}}
\newcommandx{\info}[2][1=]{\todo[linecolor=OliveGreen,backgroundcolor=OliveGreen!25,bordercolor=OliveGreen,#1]{#2}}

\newtheorem{definition}{Definition}
\newtheorem{theorem}{Theorem}

\newcommand{\ten}[1]{\ensuremath{\mathbf{#1}}}
\newcommand{\dbar}[1]{\bar{\bar{#1}}}

\newcommand{\code}[1]{\lstinline{#1}}
\newcommand{\codek}[1]{\lstinline[keywordstyle=\color{black}]{#1}}

\usepackage{lineno}
\usepackage{tikz}
\usetikzlibrary{arrows.meta, patterns}
\usetikzlibrary{shapes, calc,quotes,angles}
\tikzset{%
  >={Latex[width=2mm,length=2mm]},
  base/.style = {
    rectangle, rounded corners, draw=black, minimum width=4cm, minimum
    height=1cm, text centered, font=\sffamily},
  decision/.style = {
    diamond, draw, rounded corners, fill=blue!20, minimum width=3cm, minimum height=0.5cm, text centered, font=\sffamily},
  activityStarts/.style = {base, fill=blue!30},
  startstop/.style = {base, fill=red!30},
  activityRuns/.style = {base, fill=green!30},
  process/.style = {
    base, minimum width=2.5cm, fill=orange!15, font=\ttfamily},
  state/.style={draw, circle, minimum size=0.8cm, inner sep=0pt},
  arrow/.style={-Stealth, shorten >=1pt},
  dot/.style={minimum size=4pt, inner sep=0pt, rounded corners=1pt, fill},
}
\journal{}
\begin{document}

% set showimages to 0 to shutoff all the images
\def\showimages{1}
%%%%%%%%%%%%%%%%%%%%%%%%%%%%%%5

\begin{frontmatter}

  \title{Learning Domain-Independent Green's Function For Elliptic Partial Differential Equations}

  \author[IIT]{Pawan Negi}
  \author[IIT]{Maggie Cheng}
  \author[IITE]{Mahesh Krishnamurthy}
  \author[SOM]{Wenjun Ying}
   \author[IIT]{Shuwang Li\corref{cor1}}
  \ead{sli15@iit.edu}
\address[IIT]{Department of Applied Mathematics, Illinois Institute of Technology, Chicago, USA}
\address[IITE]{Department of Electrical and Computer Engineering, Illinois Institute of Technology, Chicago, USA}
\address[SOM]{School of Mathematical Sciences, MOE-LSC and Institute of Natural Science, Shanghai Jiao Tong University, Shanghai, China}

\cortext[cor1]{Corresponding author}

\begin{abstract}

Green's function characterizes a partial differential equation (PDE) and
maps its solution in the entire domain as integrals. Finding the analytical
form of Green's function is a non-trivial exercise, especially for a PDE
defined on a complex domain or a PDE with variable coefficients. In this
paper, we propose a novel boundary integral network to learn the
domain-independent Green's function, referred to as BIN-G. We evaluate the
Green's function in the BIN-G using a radial basis function (RBF)
kernel-based neural network. We train the BIN-G by minimizing the residual
of the PDE and the mean squared errors of the solutions to the boundary
integral equations for prescribed test functions. By leveraging the
symmetry of the Green's function and controlling refinements of the RBF
kernel near the singularity of the Green function, we demonstrate that our
numerical scheme enables fast training and accurate evaluation of the
Green's function for PDEs with variable coefficients. The learned Green's
function is independent of the domain geometries, forcing terms, and
boundary conditions in the boundary integral formulation. Numerical
experiments verify the desired properties of the method and the expected
accuracy for the two-dimensional Poisson and Helmholtz equations with
variable coefficients.

\end{abstract}


\end{frontmatter}

% \linenumbers

\section{Introduction}
\label{sec:intro}

\footnote{© 2024. This manuscript version is made available under the
CC-BY-NC-ND 4.0 license
\url{https://creativecommons.org/licenses/by-nc-nd/4.0/}}
Elliptic partial differential equations (PDEs) arise in many research
areas, such as fluid dynamics, geophysics, electrostatics,
electro-magnetics, image processing, and materials science. Green's
function, also known as the fundamental solution of the PDE, enables one to
write the solution of the PDE as integrals. With the help of the Green's
function, computational methods based on integral formulations have been
developed and show excellent accuracy and efficiency
\cite{wrobel2002,crouch1983boundary,strain1989,sidi1988,hou2002convergence,zhao2023simulating,kress1995numerical,greenbaum1993laplace,mogilevskaya2001galerkin,geng2013treecode,greengard2009fast,beale2001method,duffy2015green}.
Green's function also plays a significant role in the analysis of PDEs and
helps to establish the well-posedness and regularity properties of the PDEs
\cite{evans2010partial}. Usually, both the PDE analysis and numerical
methods require an analytical form of the Green's function to establish
theories and numerical schemes (e.g., quadratures).

Recently, as a variant of the classic integral methods, the kernel-free
boundary integral method (KFBIM) was proposed
\cite{yingKernelfreeBoundaryIntegral2007,caoKernelfreeBoundaryIntegral2022}.
One of the salient features of the KFBIM is that it does not
require an explicit form of the Green's function or special quadratures to
directly evaluate integrals. The main idea behind KFBIM is to
reinterpret the boundary integrals as solutions to an equivalent simple interface
problems. It can be solved efficiently using Cartesian grid-based
methods. Compared to the traditional finite difference or finite element
methods, the KFBIM produces a well-conditioned
linear systems and requires only a fixed number of iterations to converge
(when an iterative method is applied). The KFBIM has been successful in
numerically solving elliptic PDEs in two and three dimensions
\cite{xie2023,Dong2023,zhao2023kernel,zhou2023kernel}. However, one still does not
gain insight into the Green's function (fundamental solution) of the PDE.

Methods for finding the Green’s functions include analytically deriving
formulas or computing eigenfunction expansions for PDEs defined on simple
geometries or numerically solving a singular PDE (e.g., by approximating
the Dirac delta function).  However, when the geometry of the domain is
complex, or the PDE has variable coefficients, finding the analytical form
of the Green's function is indeed a non-trivial exercise.

In recent years, neural networks (NNs) have been employed to solve partial
differential equations owing to the development of automatic
differentiation, see
\cite{baydin2018automatic,lei2020machine,lei2023machine,qiu2021cell,zhao2022two,zhao2021convergence}
and many others. Using these neural networks, the boundary value or
interface problem can be solved by simply adding a loss term for the
interface or boundary condition
\cite{karniadakisExtendedPhysicsInformedNeural2020}.
\citet{tsengCuspcapturingPINNElliptic2022} proposed to introduce a
cusp-enforced level set function as an additional feature to the NN.
However, the physics-informed neural networks (PINNs) approach is known to
have poor convergence for boundary value problem
\cite{shengPFNNPenaltyfreeNeural2021,linBINetLearningSolve2021,sukumarExactImpositionBoundary2022}.

In order to overcome this issue, researchers proposed boundary
integral-based neural networks. \citet{linBINetLearningSolve2021} used
known Green's functions to evaluate solutions to various boundary value
problems. They employed single and double-layer potentials as the loss
function. \citet{linBIGreenNetLearningGreen2022} extended the approach to
approximate the Green's function for the boundary value problem, provided
that the Green's function for the infinite space is available.
\citet{boulleDatadrivenDiscoveryGreen2022} used the Gaussian process to
generate arbitrary pairs of test and source functions, which are then used
to learn the Green function as well as the homogeneous solution of the
given PDE. However, the learned Green's functions were specific to a
particular domain and boundary condition. Very recently,
\citet{tengLearningGreenFunctions2022} proposed to learn Green's function
for the PDEs with a given domain by approximating the Dirac delta with a
Gaussian function, which allows one to readily employ the learned Green's
function for this domain with any other set of boundary conditions.
\citet{pengDeepGeneralizedGreen2023} replaced the Dirac delta with an
alternative input for which an analytical form can be calculated for a
given boundary condition on an arbitrarily shaped domain. However, one must
solve the network again for the given PDE defined in a different domain.

In this work, we propose a novel boundary integral based neural network
that employs a radial basis function (RBF) kernel-based neural network to
learn domain-agnostic Green's function of an elliptical PDE. Since the
learned Green's function is not limited to a specific domain, it can be
readily employed in the integral formulations for solving a moving
interface or a boundary problem in fluids, materials, and wave propagation
\cite{zhao2017efficient,liu2017dynamics,kress1995numerical,xie2022fourth,pham2018nonlinear,feng2014parallel}.
Especially, problems defined in a heterogeneous media, where PDE
coefficients are spatially dependent.

We construct boundary integral neural network with a trainable Green's
function referred to as BIN-G to learn solutions of an elliptic PDE. In
BIN-G, we use single and double-layer potentials to satisfy boundary
conditions. We evaluate the density function in these potentials using
multi-layer perceptron (MLP) networks. We train the BIN-G by minimizing the
residual of the PDE such that the Green function is the solution, together
with the mean-squared error in the solution using the boundary integral
equation for the prescribed test functions. To gain efficiency, we exploit
the symmetry property of Green's function in our network, allowing us to
learn the Green's function using a 1D sample space. Furthermore, the
boundary integral formulation allows one to learn the density functions
using boundary samples alone, resulting in further reduction in the
computational cost.

Our salient contributions are as follows:
\begin{itemize}
\item We construct a RBF kernel-based neural network that allows us to
preset the approximation points and their support radius near the
singularity. This offers faster and more accurate learning compared to the
MLP networks.
\item We propose loss function that includes both the loss due to the PDE
and the  boundary integral equations. It offers learning the
domain-independent Green's function of the PDE and the domain-dependent
density functions during the training process.
\item We employ two distinct sample spaces over which the losses are
computed. A relatively large domain on which the residual of the PDE is
minimized, allows one to employ the learned domain-independent Green's
function to larger domains.
\item The learned Green's function can be readily applied to train new
density functions on arbitrarily-shaped domains and sets of boundary
conditions.
\end{itemize}
We perform numerical experiments that verify the desired properties of the
methods and the expected accuracy for the two-dimensional Poisson equation
and Helmholtz equation with variable coefficients.

In the next section, we present mathematical preliminaries related to the
boundary integral formulation. In \cref{sec:pinn}, we present the
architecture of the proposed neural network. In \cref{sec:bing}, we discuss
the training strategy to learn the domain-independent Green's function. In
\cref{sec:den_train}, we discuss the generalization of the learned Green's
function to different domain and boundary conditions. In
\cref{sec:results}, we demonstrate the capability of the neural network to
learn known Green's functions followed by learning the Green's function of
variable coefficient PDEs. In \cref{sec:conclusions}, we summarize and
discuss the outcome of the present study and future work.

\section{Mathematical Preliminaries}
\label{sec:kfbim}

The boundary integral method is widely used to solve elliptic boundary
value problems. Consider a scalar field $u$ defined in a domain $\Omega$
with boundary $\partial \Omega$, an elliptic partial differential equation
takes the form
\begin{equation}
    \mathscr{L}_\ten{x} u(\ten{x}) = f(\ten{x}), \ \forall \  \ten{x} \in \Omega,
    \label{eq:pde}
\end{equation}
where $\mathscr{L}_\ten{x}$ is an elliptic differential operator and
$f(\ten{x})$ is a source (forcing) function. We use the notation
$\mathscr{L}_\ten{x}$ to specify the variable on which the operation is
being performed. The above equation may be subjected to  a Dirichlet
boundary condition
\begin{equation}
    u(\ten{x}) = g^D(\ten{x}),  \  \forall \  \ten{x} \in \partial \Omega,
    \label{eq:dir_bc}
\end{equation}
or a Neumann boundary condition
\begin{equation}
    \frac{\partial u(\ten{x})}{\partial \ten{\nu}} = g^N(\ten{x}),  \ \forall \  \ten{x} \in \partial \Omega,
    \label{eq:neu_bc}
\end{equation}
where $\ten{\nu}$ is the outward normal to the boundary. The boundary may also be
subjected to Dirichlet on the part of the boundary and Neumann on the rest
of the boundary. A problem defined using \cref{eq:pde} and \cref{eq:dir_bc}
is known as an interior Dirichlet problem. Similarly, a problem defined
using \cref{eq:pde} and \cref{eq:neu_bc} is known as an interior Neumann
problem. A similar kind of problem may be defined for exterior domains and
interfaces \cite{kellogg1953foundations}.

The fundamental solution $G(\ten{x}, \ten{y})$ of the PDE in \cref{eq:pde}
satisfies
\begin{equation}
  \mathscr{L}_\ten{x} G(\ten{x}, \ten{y}) = -\delta(\ten{x}, \ten{y}), \ \forall \  \ten{x} \in \Omega,
  \label{eq:gf_pde}
\end{equation}
where $\ten{y}$ is the center of the function $G$. The fundamental solution
is the Green's function on an infinite domain such that it satisfies the
Sommerfeld radiation condition
\begin{equation}
  \left(\frac{\ten{x}}{|\ten{x}|}, \nabla u(\ten{x}) \right) - i \kappa
  u(\ten{x}) = o\left(\frac{1}{|\ten{x}|}\right), |\ten{x}| \to \infty,
\end{equation}
where $\kappa = 0$ in the present case. The solution of \cref{eq:pde} with
homogeneous Dirichlet boundary condition is given by
\begin{equation}
  u(\ten{x}) = -\int_\Omega f(\ten{y}) G(\ten{x}, \ten{y}) d\ten{y},
  \label{eq:gf_f}
\end{equation}
where $G(\ten{x}, \ten{y})$ also satisfies the homogeneous boundary
condition i.e $G(\ten{x}, \ten{y})=0, \ \forall \ \ten{x} \in
\partial \Omega$ along with \cref{eq:gf_pde}. In the case of the
non-homogeneous boundary conditions, we use the following definitions and
theorems from the potential theory \cite{kellogg1953foundations}.
\begin{definition}
Let $g$ be a continuous function defined on $\partial \Omega$. The single layer
potential
\begin{equation}
  \bar{u}(\ten{x}) := - \int_{\partial \Omega} g(\ten{y}) G(\ten{x},
  \ten{y}) dS(\ten{y}).
\end{equation}
Similarly, the double-layer potential
\begin{equation}
  \dbar{u}(\ten{x}) := - \int_{\partial \Omega} h(\ten{y}) \frac{\partial
  G}{\partial \ten{\nu}_y}(\ten{x}, \ten{y}) dS(\ten{y}),
\end{equation}
where $h$ is a continuous function defined on $\partial \Omega$,
$\ten{\nu}_y$ is the outward normal to the boundary at $\ten{y}$.
\label{def:single_double}
\end{definition}

\begin{theorem}
  Given a PDE of the form $\mathscr{L}_\ten{x} u(\ten{x}) = f(\ten{x}), \
\forall \  \ten{x} \in \Omega$ with a non-homogenous boundary condition, the
solution of the interior Dirichlet problem is given by
\begin{equation}
  u(\ten{x}) = \int_\Omega f(\ten{y}) G(\ten{x}, \ten{y}) d\ten{y} + \dbar{u}
  \label{eq:gf_sol_dir}
\end{equation}
with
\begin{equation}
  \lim_{\ten{x} \to \ten{x}_o^-} u(\ten{x}) = -\frac{1}{2} h(\ten{x_o}) +
  u(\ten{x_o}), \ \forall \ \ten{x}_o \in \partial \Omega.
  \label{eq:dir_bc_cond}
\end{equation}
where $\ten{x}_o^-$ mean converging in the interior of $\Omega$. Similarly,
for the interior Neumann problem, the solution is given by
\begin{equation}
  u(\ten{x}) = \int_\Omega f(\ten{y}) G(\ten{x}, \ten{y}) d\ten{y}
   + \bar{u}
   \label{eq:gf_sol_neu}
\end{equation}
with
\begin{equation}
  \lim_{\ten{x} \to \ten{x}_o} u(\ten{x}) = u(\ten{x_o})  \ \forall \ \ten{x}_o \in \partial \Omega.
\end{equation}
\label{thm:sols}
\end{theorem}
We note that the $G(\ten{x}, \ten{y})$ in \cref{thm:sols} satisfies the
homogenous boundary condition on $\partial \Omega$ along with
\cref{eq:gf_pde}. However, one can also evaluate the solution using both
the first layer and second layer potential
\begin{equation}
    u(\ten{x}) = - \int_\Omega f(\ten{y}) G(\ten{x}, \ten{y}) d\ten{y} + \dbar{u} - \bar{u},
    \label{eq:main_eq}
\end{equation}
where $G$ is a domain independent Green's function that does not require to
satisfy homogeneous boundary conditions. Furthermore, at the boundary, we
use the condition in \cref{eq:dir_bc_cond} irrespective of the type of the
boundary condition.

Traditional integral approaches require the analytical form of the Green's
function to design quadratures. However, it is challenging to calculate an
explicit expression for the Green's function, especially when the PDEs are defined
in a complex domain or have variable coefficients. To circumvent this
constraint, a kernel-free boundary integral methods have been proposed
\cite{yingKernelfreeBoundaryIntegral2007,caoKernelfreeBoundaryIntegral2022}
and are widely used to solve variable coefficient PDEs. However, this
method does not produce any information about Green's function of the PDE.
In the next section, we develop a neural network-based approach to tackle
the problem.

\section{Boundary integral network architecture}
\label{sec:pinn}

Recently, \citet{linBINetLearningSolve2021} proposed Boundary integral
Network (BINet), where a known domain independent Green's function of the
PDE is used, and the density function $h$ or $g$ in the
\cref{def:single_double} is learned using neural networks. In our method,
we do not use the analytical Green's function and use three distinct
networks to learn $G(\ten{x}, \ten{y})$, $h(\ten{x})$, and $g(\ten{x})$
functions using test functions. We call the proposed \textbf{B}oundary
\textbf{I}ntegral \textbf{N}etwrok with unknown \textbf{G}reen's function
as BIN-G. We use $G(\ten{x}, \ten{y}) \equiv G(\ten{x}, \ten{x}_c)$, where
$\ten{x}_c$ is the center of the Green's function in the following
discussion.

\begin{figure}[!htpb]
\centering
\scalebox{0.85}{
\begin{tikzpicture}
  \node[] (x1) {$\ten{x}$};
  \node[below =of x1, yshift=-2.5cm] (xc1) {$\ten{x}^b_1$};
  \node[below=of xc1, yshift=-0.25cm] (xc2) {$\ten{x}^b_n$};
  \node[dot] at ($(xc1)!0.33!(xc2)$) {};
  \node[dot] at ($(xc1)!0.5!(xc2)$) {};
  \node[dot] at ($(xc1)!0.66!(xc2)$) {};
  \node[state, right=of x1] (minus1) {$-$};
  \node[state, below=of minus1, yshift=-0.8] (minus2) {$-$};
  \node[right=of minus1, xshift=-0.45cm, yshift=0.3cm] (r1) {$r_1$};
  \node[below =of r1, yshift=-0.3cm] (r2) {$r_n$};
  \node[dot] at ($(minus1)!0.33!(minus2)$) {};
  \node[dot] at ($(minus1)!0.5!(minus2)$) {};
  \node[dot] at ($(minus1)!0.66!(minus2)$) {};
  \node[base, right=of minus1, xshift=1cm, fill=orange] (g1) {$-b_1\text{KNN}_G(r_1; \Theta)$};
  \node[base, right=of minus2, xshift=1cm, fill=orange] (g2) {$-b_n\text{KNN}_G(r_n; \Theta)$};
  \node[dot] at ($(g1)!0.33!(g2)$) {};
  \node[dot] at ($(g1)!0.5!(g2)$) {};
  \node[dot] at ($(g1)!0.66!(g2)$) {};
  \node[base, right=of xc1, xshift=2.7cm] (h1) {$\text{MLP}_g(\ten{x}_1^b; \theta_g)$};
  \node[base, below=of h1] (h2) {$\text{MLP}_g(\ten{x}_n^b; \theta_g)$};
  \node[dot] at ($(h1)!0.33!(h2)$) {};
  \node[dot] at ($(h1)!0.5!(h2)$) {};
  \node[dot] at ($(h1)!0.66!(h2)$) {};
  \node[state, right=of g1, yshift=-2.cm] (prod1) {$\times$};
  \node[state, below=of prod1] (prod2) {$\times$};
  \node[dot] at ($(prod1)!0.33!(prod2)$) {};
  \node[dot] at ($(prod1)!0.5!(prod2)$) {};
  \node[dot] at ($(prod1)!0.66!(prod2)$) {};
  \node[state, right=of prod1, yshift=-1cm] (int1) {$\sum$};
  \node[right=of int1] (out) {$u(\ten{x})$};
  \draw[arrow] (x1) -- (minus1);
  \draw[arrow] (xc1) -- (minus1);
  \draw[arrow] (x1) -- (minus2);
  \draw[arrow] (xc2) -- (minus2);
  \draw[arrow] (minus1) -- (g1);
  \draw[arrow] (minus2) -- (g2);
  \draw[arrow] (xc1) -- (h1.west);
  \draw[arrow] (xc2) -- (h2);
  \draw[arrow] (g1.east) -- (prod1);
  \draw[arrow] (h1.east) -- (prod1);
  \draw[arrow] (g2.east) -- (prod2);
  \draw[arrow] (h2.east) -- (prod2);
  \draw[arrow] (prod1) -- (int1);
  \draw[arrow] (prod2) -- (int1);
  \draw[arrow] (int1) -- (out);
\end{tikzpicture}
}
\caption{Network architecture of the BIN-G. Using the distance $r_i$ between $\ten{x}$ and $\ten{x}_i^b$ enforcing the symmetry of the Green's function. The block highlighted in orange evaluates a domain-independent Green's function learned using a KNN.}
\label{fig:kfbi_nn}
\end{figure}
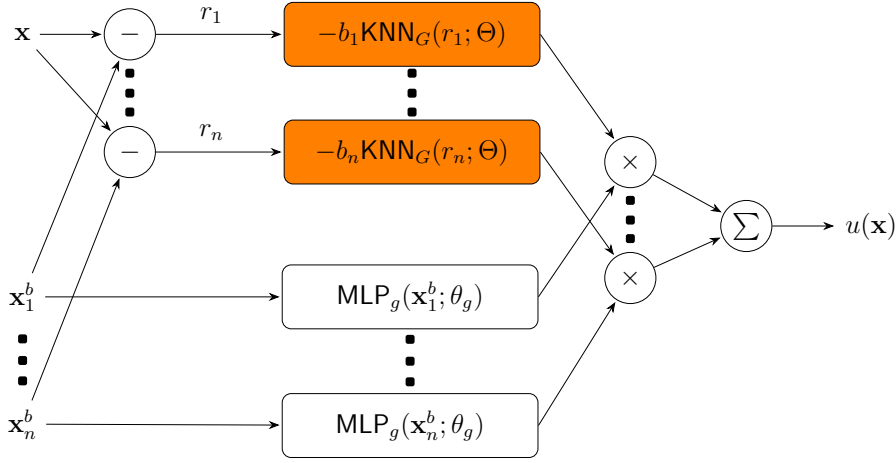

We consider an interior Neumann boundary value problem as shown in
\cref{eq:pde} with $f(\ten{x})=0$ to simply the network architecture. In
\cref{fig:kfbi_nn}, we show a typical network that evaluates the
single-layer potential, which is the solution to the present problem. It
takes the position as an input $\ten{x}$ and produces the value of the field
$u(\ten{x})$ as an output. We generate the coordinates $\{\ten{x}^b=(x^b,
y^b) | \ten{x}^b \in \partial \Omega\}$ by discretizing the boundary of
interest using $N_b$ points. The various layers involved in the network are shown as follows:
%\begin{sloppypar}
\begin{enumerate}
\item The input are coordinates $\ten{x} \in \Omega$ at which one desire to
obtained the solution.
\item The distance is calculated from all the boundary points
$\{\ten{x}_1^b, \dots, \ten{x}_n^b \}$ as an intermediate step in $r$. We
perform this step to ensure the symmetry of the Green's function while
training. One can ensure symmetry by employing a loss function that minimizes the interchanges of argument of the Green's function as well. Therefore, removing this step only slows down learning and requires
additional terms in the loss function.
\item The boundary points are fixed. We input the boundary points to the
multi-layer perceptron (MLP) network denoted by $\text{MLP}_g$ to evaluate
the density function $g$ in the \cref{def:single_double}.
\item From the distance values $r$, the Green's function value is calculated using
a kernel-based neural network (KNN) denoted by KNN$_G$ discussed in the next
section.
\item In the next two steps, numerical quadrature is performed using the
evaluated Green's and density function. We use a three-point linear element and a four-point triangular element to perform numerical
integration on the boundary and the volume, respectively.
\end{enumerate}
%\end{sloppypar}
The mathematical expression for the network shown in \cref{fig:kfbi_nn} is
\begin{equation}
  u(\ten{x}) = -\sum_{i}^{N_b} b_i \
  \text{MLP}_g(\ten{x}_i^b; \theta_g)\  \text{KNN}_G(||\ten{x} - \ten{x}^b_i||;
  \Theta),
\end{equation}
where $b_i$ are the integration weights, $\ten{x}^b_i  \in \partial \Omega$
and $\theta_g$ and $\Theta$ are the learnable parameters.

In the case of the Dirichlet boundary value
problem, we use automatic differentiation to evaluate the gradient of the
Green's function in the orange block. The mathematical expression of this
kind of network is given by
\begin{equation}
  u(\ten{x}) = -\sum_{i}^{N_b} c_i \
  \text{MLP}_h(\ten{x}_i^b; \theta_h)\  \ten{n}_i \cdot \nabla
  \text{KNN}_G(||\ten{x} - \ten{x}^b_i||; \Theta),
\end{equation}
where $c_i$ are the integration weights, $\theta_h$ are learnable
parameters, $n_i$ are the outward normal at $\ten{x}_i^b  \in \partial
\Omega$, and the gradient is evaluated using automatic differentiation.

Similarly, in the presence of a non-zero forcing function $f(\ten{x})$, we
add the volume integration $\int_\Omega G(\ten{x}, \ten{x}_c) f(\ten{x}_c)
d\ten{x}_c$ after computing boundary integral. We discretize the volume using $N_i$ quadrature points. The mathematical expression
of this kind of network is given by
\begin{equation}
  \begin{split}
  u(\ten{x}) & = \sum_{i}^{N_i} a_i f(\ten{x}_i) \text{KNN}_G(||\ten{x}
  - \ten{x}_i||; \Theta) \\
  ~& -  \sum_{i}^{N_b} b_i \ \text{MLP}_h(\ten{x}_i^b;
  \theta_h)\  \ten{n}_i \cdot \nabla \text{KNN}_G(||\ten{x} - \ten{x}^b_i||;
  \Theta)\\
  ~&- \sum_{i}^{N_b} c_i  \text{MLP}_g(\ten{x}_i^b; \theta_g)
  \text{KNN}_G(||\ten{x} - \ten{x}^b_i||; \Theta),
  \end{split}
  \label{eq:bie_main}
\end{equation}
where $a_i$ are the integration weights, $\ten{x}_i \in \Omega$. Since we
focus on learning a domain agnostic Green's function, we use the
architecture defined in \cref{eq:bie_main} following \cref{eq:main_eq} for
all our test cases. In the next section, we discuss the Green's function
network architecture.

\subsubsection{Green's function network architecture}
\label{sec:green_nn}

\begin{figure}[htpb]
\centering
\begin{tikzpicture}
  \node[] (x1) {$\ten{x}$};
  \node[below =of x1, yshift=0.3cm] (xc1) {$\ten{x}_c$};
   \node[state, right=of x1, yshift=0cm] (minus1) {$-$};
  \node[right= of minus1, yshift=0.2cm] (r1) {$r$};
  \node[right= of minus1, yshift=-1.1cm] (r2) {$r$};
  \node[base, right= of minus1, xshift=1cm] (k1) {$w_1 K(r; \zeta_1, \lambda_1)$};
  \node[base, below= of k1] (k2) {$w_n K(r; \zeta_n, \lambda_n)$};
   \node[dot] at ($(k1)!0.33!(k2)$) {};
  \node[dot] at ($(k1)!0.5!(k2)$) {};
  \node[dot] at ($(k1)!0.66!(k2)$) {};
  \node[state, right=of k1] (sum) {$\sum$};
  \node[right=of sum] (out) {$G(\ten{x}, \ten{x}_c)$};
  \draw[arrow] (x1) -- (minus1);
  \draw[arrow] (xc1) -- (minus1);
  \draw[arrow] (minus1) -- (k1.west);
  \draw[arrow] (minus1) -- (k2.west);
  \draw[arrow] (k1.east) -- (sum);
  \draw[arrow] (k2.east) -- (sum);
  \draw[arrow] (sum) -- (out);
\end{tikzpicture}
\caption{KNN-based architecture to evaluate domain-independent Green's function.}
\label{fig:spinn_green}
\end{figure}
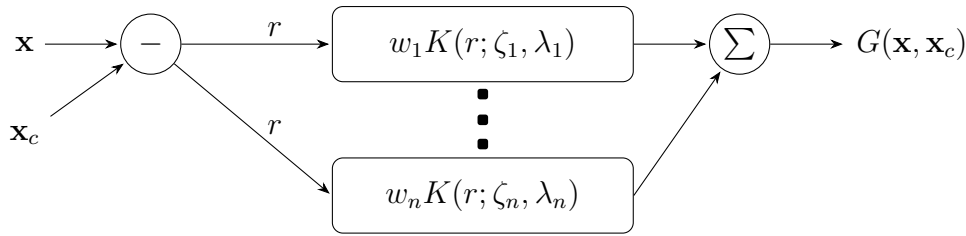

In order to evaluate Green's function for the elliptic operator
$\mathscr{L}_\ten{x}$, we use a radial basis function (RBF) kernel-based
neural network (KNN)
\cite{eMachineLearningContinuous2020a,ramabathiranSPINNSparsePhysicsbased2021}.
Any scalar function $f$ defined in a domain $\Omega$ can be approximated
using a RBF kernel $K$ as
\begin{equation}
  f(\ten{x}_i) = \sum_j w_j K(\ten{x}_i; \ten{\zeta}_j, \lambda_j),
\end{equation}
where $w_j$ is the integration weights, $\ten{\zeta}_j$ is the kernel
center, and $\lambda_j$ is the scaling parameter. Therefore, we can
approximate Green's function as $G(\ten{x}, \ten{x}_c; \Theta) = \text{KNN}_G(r_i;
\Theta)$, where
\begin{equation}
\text{KNN}_G(r_i; \Theta) = \sum_j w_j K(r_i; \zeta_j, \lambda_j),
\label{eq:gf_sph}
\end{equation}
%\begin{sloppypar}
where $r_i = || \ten{x}_i - \ten{x}_c ||$, and $\Theta=\{w_1, w_2, ... ,
w_n, \zeta_1, \zeta_2, ..., \zeta_n, \lambda_1, \lambda_2, .., \lambda_n\}$
are the learnable parameters. \Cref{eq:gf_sph} ensures that $G(\ten{x},
\ten{x}_c; \Theta) = G(\ten{x}_c, \ten{x}; \Theta)$. We note that $n$ is a
hyper-parameter denoting the number of approximating points (or particles).
In \cref{fig:spinn_green}, we show the architecture of the proposed
network. Since we convert all coordinates into the distances $r$ from the
center of the Green's function, we use a one-dimensional kernel
\begin{equation}
K(r_i; \zeta_i, \lambda_i) = \exp\left(-0.5
\left( \frac{r_i - \zeta_i}{\lambda_i}\right)^2\right).
\end{equation}
Since, the parameters $\lambda$ and $\zeta$ has numerical origin, we can
create a set of values such that a high slope function near singularity can
be learn fast. We create $n$ particles, exponential spaced in $[0, 3.0]$
with a linear increase of support radius in $[0.001, 0.2]$. The particles
near zero are closely spaced with the lowest support radius. In the next
section, we discuss the training process of all the networks in detail.
%\end{sloppypar}

\section{BIN-G training methodology}
\label{sec:bing}

In this section, we discuss the method used to train BIN-G to learn the
domain-independent Green's function for a given elliptic PDE. As discussed in
\cref{sec:kfbim}, the Green's function satisfies \cref{eq:gf_pde}. Therefore, it is
logical to minimize the following loss,
\begin{equation}
   \min\limits_{\Theta}  ||\mathscr{L}_\ten{x} G(\ten{x}, \ten{x}_c; \Theta) + \tilde{\delta}(\ten{x} - \ten{x}_c) ||^2_\Omega,
  \label{eq:gf_loss}
\end{equation}
where $\tilde{\delta}$ is an approximation of the Dirac delta function.
\citet{tengLearningGreenFunctions2022} proposed to use a Gaussian
approximation for the Dirac delta and train the NN to learn the solution
like a traditional MLP network. However, \citet{linBINetLearningSolve2021}
showed that traditional MLPs are unable to converge for the Helmholtz
equation. In other words, the presence of Gaussian approximation prevents
the network from learning the desired slope near the singularity. In this
paper, we mitigate this issue by learning the slopes near singularity using
boundary integral formulation.

\subsection{Green's function training scheme}
\label{sec:gf_train}

\begin{figure}
  \centering
  \begin{subfigure}[b]{0.48\textwidth}
      \centering
      \includegraphics[width=\textwidth]{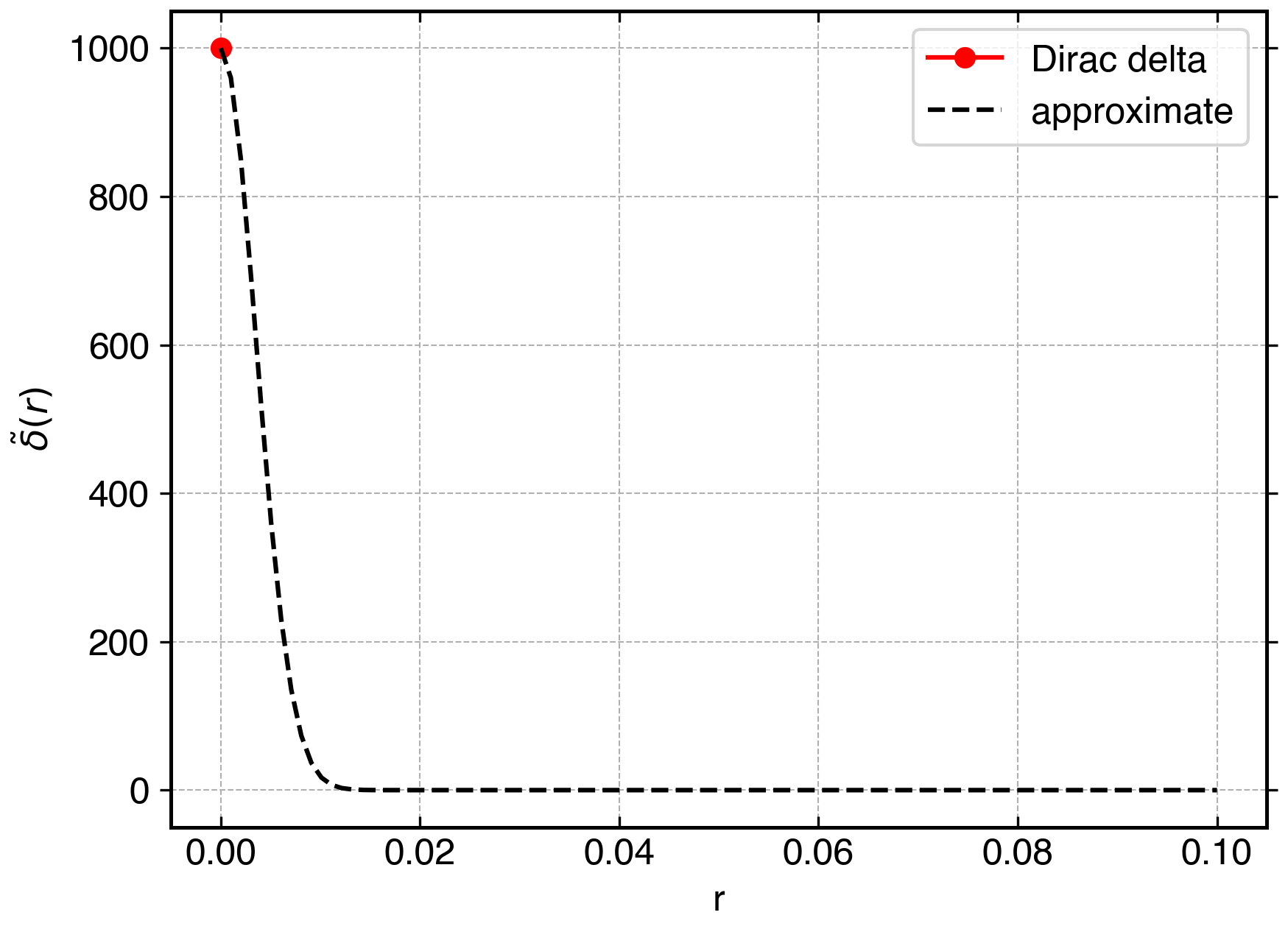}
      \caption{Dirac delta function}
      \label{fig:dirac_delta}
  \end{subfigure}
  \hfill
  \begin{subfigure}[b]{0.48\textwidth}
      \centering
      \includegraphics[width=\textwidth]{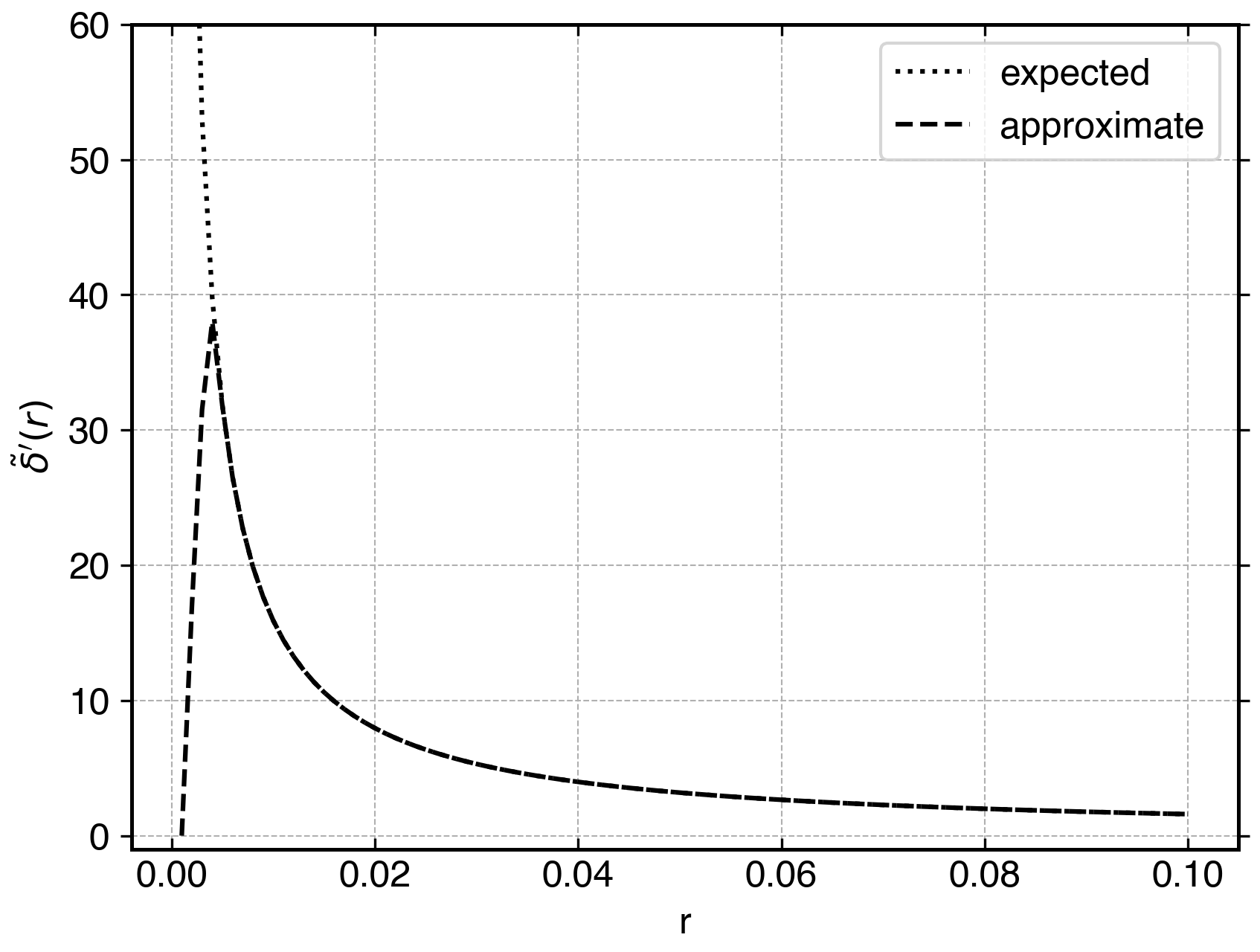}
      \caption{Derivative of a typical Green's function}
      \label{fig:gf_derv}
  \end{subfigure}
\caption{Comparison of the approximation of the Dirac delta and the
expected derivative of the Green's function at the singularity.}
     \label{fig:approximated_func}
\end{figure}

We note that the NN cannot learn an infinite slope. Furthermore, the slope
of the learned Green's function will  be zero at the singularity due to
continuity assured by the NNs. In \cref{fig:approximated_func}, we show a
schematic plot for the Dirac delta approximation and the derivative of the
Green's function. Therefore, it is very difficult to find an approximation
of the Dirac delta such that it does not influence the Green's function,
has a zero slope at the singularity, and has a very large slope close to
the singularity. In order to remedy this problem, we satisfy the PDE away
from singularity on a 1D domain for a large length. Additionally, we solve
a boundary value problem for two distinct test functions on a finite
domain.

\begin{figure}[htbp]
  \centering
  \includegraphics[width=0.5\linewidth]{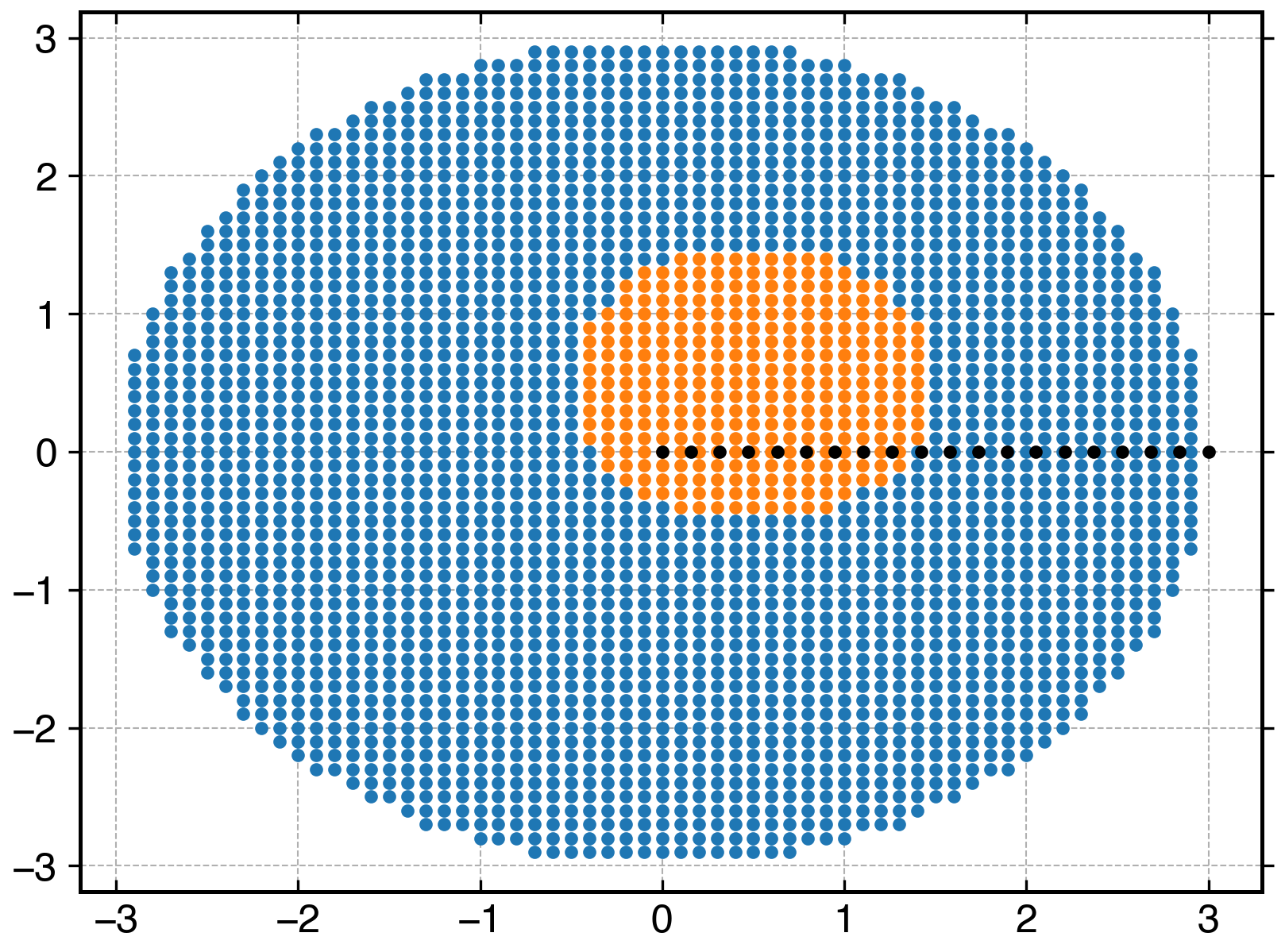}
  \caption{The sample space used in the training of the domain independent Green's
  function network. The domain used to compute the loss due to the PDE is a 1D
domain shown in black, representing the domain in blue for a 2D PDE (due to symmetry of the Green's function). The orange domain is used to compute loss due to boundary integral equations.}
  \label{fig:gf_train_domain}
\end{figure}

In the \cref{fig:gf_train_domain}, we show different sample spaces used in
the training of the Green's function. In the sample space shown in blue, we
obtain the loss due to the PDE. Since the Green's function learned are
radially symmetric, we can also obtain samples from a 1D domain shown in
black color. Therefore, we obtain samples from a one-dimensional domain
$\Omega_{1D}:= [0, 3]$ to evaluate the residue of the PDE. However, we
ignore samples in $S_\alpha = \{ x | x - x_c < \alpha, x \in
\Omega_{1D}\}$, where $\alpha$ is a hyper-parameter set to $0.01$. We
assume $x_c=0.0$ in the 1D domain. The loss due to the PDE
\begin{equation}
  \mathcal{L}_{PDE} = \left| \mathscr{L}_{\ten{x}} \text{KNN}_G(\ten{x}, \ten{x}_c; \Theta) \right|^2,
\end{equation}
where $\ten{x} \in \Omega_{1D} - S_\alpha$, $\Theta$ are the learnable
parameters for the KNN. The function $|\bullet|$ is a discrete $L_2$ norm
given by
\begin{equation}
    \sqrt{\sum_i^{N} \frac{(\bullet)^2}{N}},
\end{equation}
where $N$ is the number of samples.

In a 2D domain, much smaller than the blue domain, we evaluate the solution
using the BIN-G as discussed in \cref{sec:pinn}. We use the test functions
$\phi_1(\ten{x} \equiv (x, y)) = \sin(2 \pi x) \sin(2 \pi y)$ and
$\phi_2(\ten{x}\equiv(x, y)) = \exp(-(x^2 + 2 y^2 + 1))$ in the domain
$\Omega$ in orange in \cref{fig:gf_train_domain}. The boundary condition
and the forcing function can be obtained for both the test function using
the domain information and the PDE. We employ a circular domain with radius
$0.5$ centered at $(0.5, 0.5)$. We sample $N$ random points from the
interior of the $\Omega$ and compute the mean squared error for prescribed
test functions. The loss due to test function $\phi_1$
\begin{equation}
  \mathcal{L}_{BI_1} = \left|(u(\ten{x}; \theta, \Theta) - \phi_1(\ten{x})\right|^2,
\end{equation}
and due to test function $\phi_2$
\begin{equation}
  \mathcal{L}_{BI_2} = \left|(u(\ten{x}; \theta, \Theta) - \phi_2(\ten{x})\right|^2,
\end{equation}
where $u(\ten{x}_i)$ is evaluated using the BIN-G, $\theta=\{ \theta_g,
\theta_h \}$ and $\Theta$ are
the learnable parameters for the MLP and KNN, respectively.

Combining all the component, the total loss used to train the BIN-G network is given by
\begin{equation}
  \mathcal{L}(\theta, \Theta) =  \beta \mathcal{L}_{PDE} + \gamma (\mathcal{L}_{BI_1} + \mathcal{L}_{BI_2}) + \eta \mathcal{L}_N,
\end{equation}
where $\beta=1.0$, $\gamma=[0.1, 1.0]$, and $\eta=[0.01, 0.1]$. We define
\begin{equation}
    \mathcal{L}_N = \left|\text{KNN}_G(\ten{x}, \ten{x}_c; \Theta)
    \right|^2 + \left| \text{MLP}_h(\ten{x}_b; \theta_h)\right|^2 + \left|
    \text{MLP}_g(\ten{x}_b; \theta_g)\right|^2,
\end{equation}
a loss normalization of the function magnitudes. It ensures that the
magnitude of the functions is minimal. This cannot be achieved by the usual
$L_2$ normalization where the loss due to the magnitude of the parameter
squared is minimized due to the presence of the KNN in the network, which
assumes a large value of the parameters near the singularity. We note that
the contribution of $\mathcal{L}_N$ is very low compared to other loss
functions.

We train the NNs using Adam optimizer with a learning rate of $10^{-4}$ for
the $10^5$ epochs unless stated otherwise. Since the learned Green's
function is radially symmetric, the 1D domain $\Omega_{1D}$ captures a
larger domain with a much smaller number of samples. This reduces the
training time significantly. We note that the Green's function training
simultaneously trains the density functions $h$ and $g$ as well. However,
one can readily employ the learned Green's function for any other domain
and boundary condition using the method described in the next section.

\section{Generalization to other domain and boundary conditions}
\label{sec:den_train}

In this section, we use the learned Green's function to obtain the solution to a problem with the same governing PDE but different domain and boundary conditions. We note that the Green's function is not in the exact analytical form, but rather parameterized by a neural network. Since it is a free-space Green's function, it can be employed to different domain and boundary conditions using \cref{eq:main_eq}.

In order to learn the PDE solution for a new domain and boundary conditions, we only need to learn the density functions
$\text{MLP}_h$ and MLP$_g$ in \cref{eq:bie_main}, which are parameterized by multilayer perceptron neural networks. In the following, we describe the training of
neural networks (MLP$_h$ and MLP$_g$) for a interior
Dirichlet problem. We use \cref{thm:sols} to obtain the loss function
\begin{equation}
  \mathcal{L}(\theta) =
    |g^D(\ten{x}) - (u(\ten{x}, \theta) + 0.5 h(\ten{x}, \theta))|^2,  \ \forall \ \ten{x} \in \partial \Omega,
  \label{eq:df_loss}
\end{equation}
where $\ten{x}$ is the input, $g^D$ is the given Dirichlet boundary
condition and $\theta$ is the set of trainable parameters in the MLP in Figure \ref{fig:kfbi_nn}.

\begin{sloppypar}
In algorithm \ref{alg:reuse_gf}, we show the pseudo-code for the training procedure using learned Green's function. We need samples from the boundary only. In the \codek{load_nn_network}, we load a trained KNN network which evaluated the GF for the given coordinates and center. In \codek{initialize_h_network}, and \codek{initialize_g_network}, we initialize the MLP$_h$ and MLP$_g$ network with uniformly random values. The \codek{while} loop is the traditional training process where \codek{initialize_grad} resets the gradient values to zero, and the \codek{test_function(x, y)} are solution on the boundary for an interior Dirichlet problem. The \codek{eval_sol} is the BIN-G network that utilizes the learned GF network and evaluates the PDE solution. We compute the loss in \cref{eq:df_loss} in the variable \codek{loss}, and use the Adams optimizer in \codek{optimizer_step} as was done in the learning of the Green's function.
\end{sloppypar}

 \begin{algorithm}[h!]
   \SetAlgoLined
   \KwIn{$\{\{(x, y) | x, y \in \partial \Omega\}$}
   \KwResult{trained KNN$_G$ network}
   \code{n_train} = $10^5$\;
   \codek{KNN_G = load_knn_network()}\;
   \codek{MLP_h = initialize_h_network()}\;
   \codek{MLP_g = initialize_g_network()}\;
   \While{\codek{i < n_train}} {
     \codek{initialize_grad()}\;
     \codek{u = eval_sol(x, y, KNN_G, MLP_h, MLP_g)}\;
     \codek{u_on_boundary = u + MLP_h(x, y)/2}\;
     \codek{loss = mean_squared(u_on_boundary - test_function(x, y))}\;
     \codek{optimizer_step(loss, MLP_h, MLP_g, ...)}\;
     \code{i++}\;
   }
   \caption{Pseudo-code to learn density function employing a leaned Green's function.}
   \label{alg:reuse_gf}
 \end{algorithm}

The proposed method is a straightforward application of boundary integral
formulation. However, one possible future endeavor is to express the
boundary using a parameterized curve following the method proposed by
\citet{mezzadriFrameworkPhysicsInformedDeep2023}, and train the
parameterized network using samples from different domain shapes, resulting
from different curve parameters.

\section{Results and discussion}
\label{sec:results}

In this section, we demonstrate the applicability of the proposed method to
solve various elliptic partial differential equations. We first learn the
Green's function for the Laplace equation. We then learn the Green's
function for the Helmholtz equation for different $k$ values. Finally, we
learn the Green's function of a variable coefficient elliptic PDE. For all
the test cases, we readily employ the trained Green's function network to
train the density functions MLP$_g$ and MLP$_h$ to learn the solution on a
different domain and boundary conditions. In order to verify the learned
GF, we use a new test function
\begin{equation}
u(\ten{x} \equiv (x, y)) = \exp(-x) \cos(y) + \exp(-y) \sin(x),
\label{eq:test_func}
\end{equation}
on a square-shaped domain of unit length for all the test cases unless
stated otherwise. We use the test function in \cref{eq:test_func} to obtain
the forcing function and boundary condition and then use that as prescribed
data to the network to evaluate the test function. All the NN present in
this work are implemented using the open source \texttt{pytorch}
\cite{NEURIPS2019_9015} package.

For all test cases, we consider the KNN$_{G}$ network discussed in
\cref{sec:green_nn} with $400$ approximation points, which results in
$1200$ parameters to be trained. On the testing data, for all the results,
we report relative $L_2$ error
\begin{equation}
  L_2 =\left(\frac{ \sum_i^N (f(\ten{x}_i) - f_o(\ten{x}_i))^2
  }{ \sum_i^N ( f_o(\ten{x}_i))^2
  }\right)^{\frac{1}{2}},
  \label{eq:error_l2}
\end{equation}
where $N$ is the number of test points, $f_o$ is the known test function
and $f$ is the value obtained using the BIN-G. We run all the simulations
on a ``Apple M2 Ultra Chip''. It takes $5.7$ hours to train the Green's
function for a variable coefficient PDE. However, for the constant
coefficient PDEs since we use a $1D$ sample space, it learns the GF in $3$
hours. Furthermore, for all the cases it takes approximately $9.5$ minutes
to retrain the density functions using the learned GF.

\subsection{Laplace equation}
\label{sec:lap}

\begin{figure}
  \centering
  \begin{subfigure}[b]{0.45\textwidth}
      \centering
      \includegraphics[width=\textwidth]{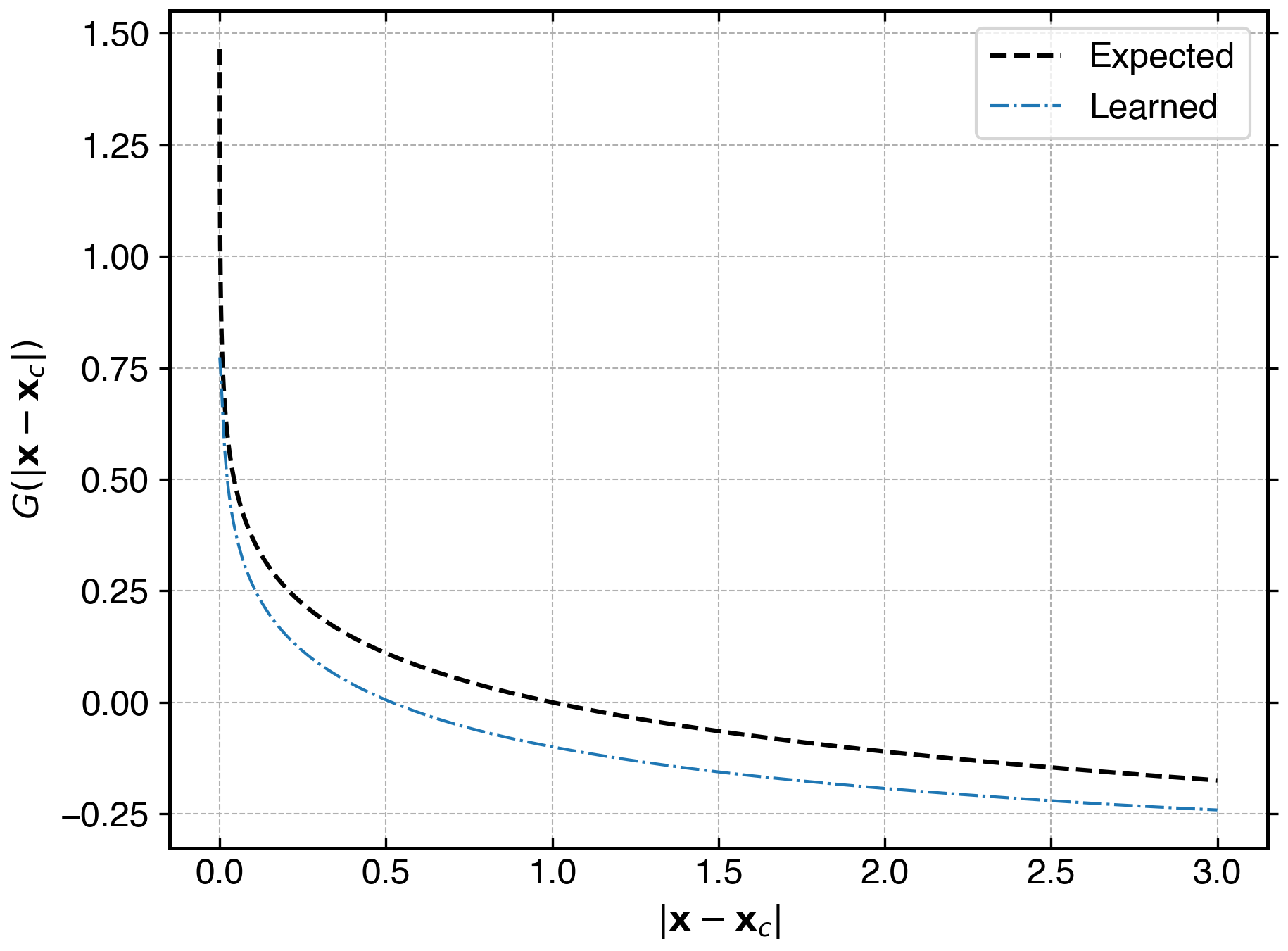}
      \caption{Green's function}
      \label{fig:lap_gf}
  \end{subfigure}
  \begin{subfigure}[b]{0.45\textwidth}
      \centering
      \includegraphics[width=\textwidth]{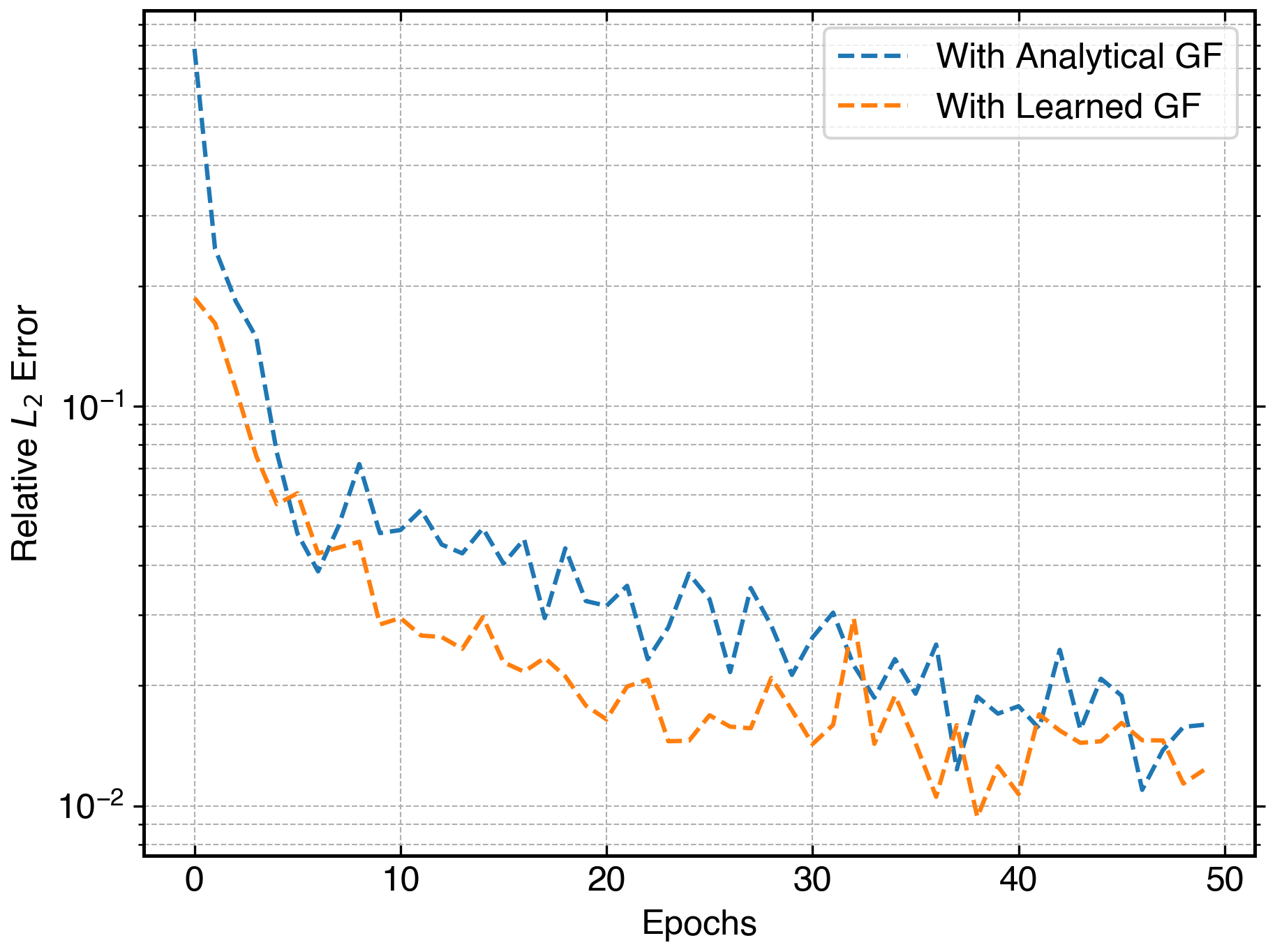}
      \caption{$L_2$ error on test data}
      \label{fig:lap_l1_error}
  \end{subfigure}
\caption{Comparison of the learned and analytical Green's functions and
convergence of $L_2$ error in the test data while training density function
using learned and analytical Green's function of Laplace equation.}
     \label{fig:lap_force}
\end{figure}

In this section, we first solve the Laplace equation. The fundamental
solution of the Laplace equation is given by $G(\ten{x}, \ten{x}_c) = 1/2
\pi \ln(|\ten{x} - \ten{x}_c|)$. We consider a test function in
\cref{eq:test_func} such that
\begin{equation}
  \nabla^2 u(\ten{x}) = f(\ten{x}),
  \label{eq:forcing_lap}
\end{equation}
where $f(x) = 0$. In \cref{fig:lap_gf}, we plot the learned Green's
function against the analytical result. The learned Green's function is
slightly shifted. However, the shift does not affect the solution in the
case of the Laplace equation. We employ the learned Green's function to
train the density function networks MLP$_h$ and MLP$_g$ for the test
function in \cref{eq:test_func}.
In \cref{fig:lap_l1_error}, we show the $L_2$ error with the number of
epochs showing convergence when an analytical and a learned Green's function is employed
in the BIN-G. The error in the test data shows a similar trend for a
learned Green's function compared to an analytical Green's function.

\begin{figure}
  \centering
  \begin{subfigure}[b]{0.45\textwidth}
      \centering
      \includegraphics[width=\textwidth]{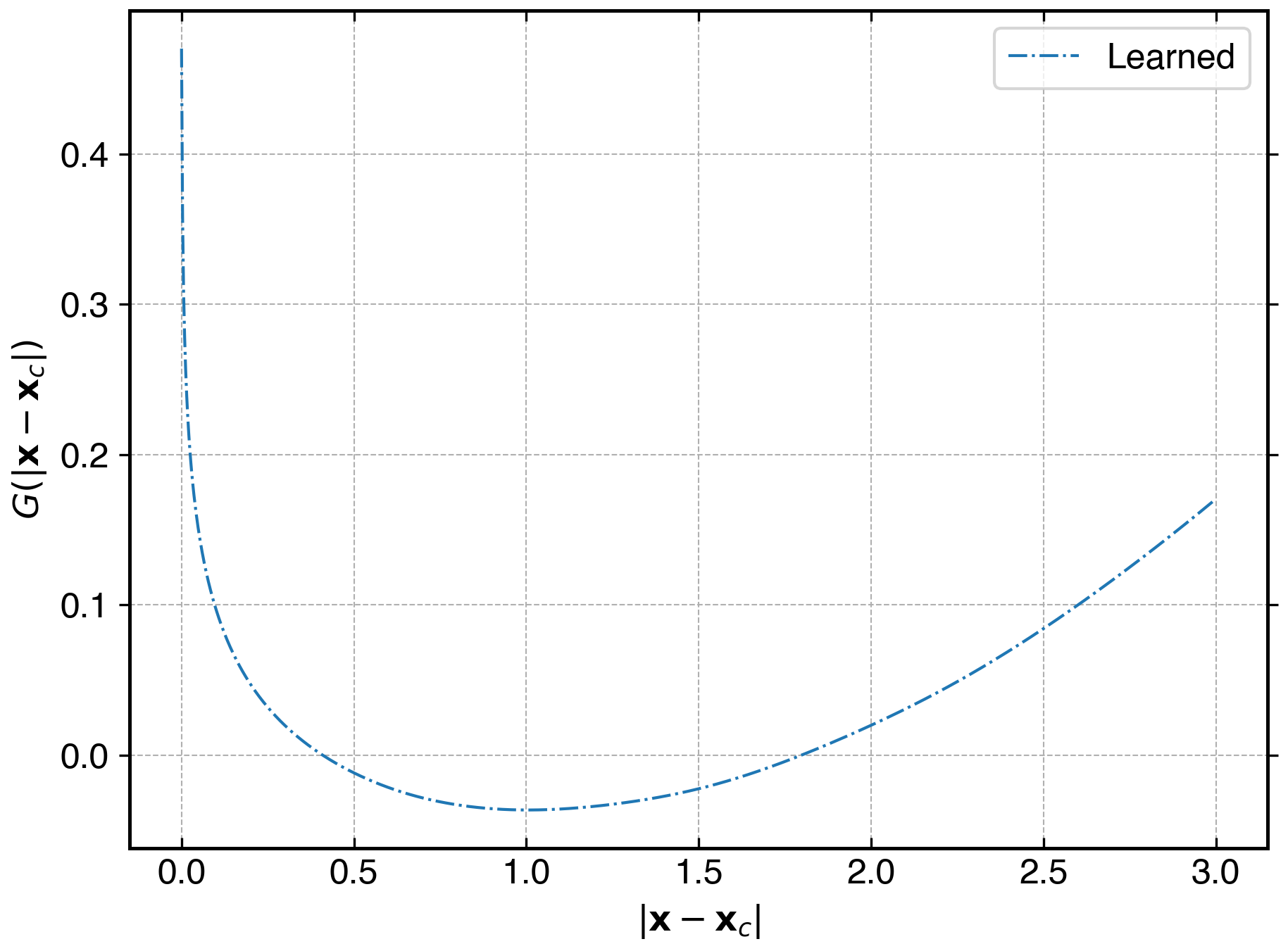}
      \caption{Green's function}
      \label{fig:var_lap_gf}
  \end{subfigure}
  \begin{subfigure}[b]{0.45\textwidth}
      \centering
      \includegraphics[width=\textwidth]{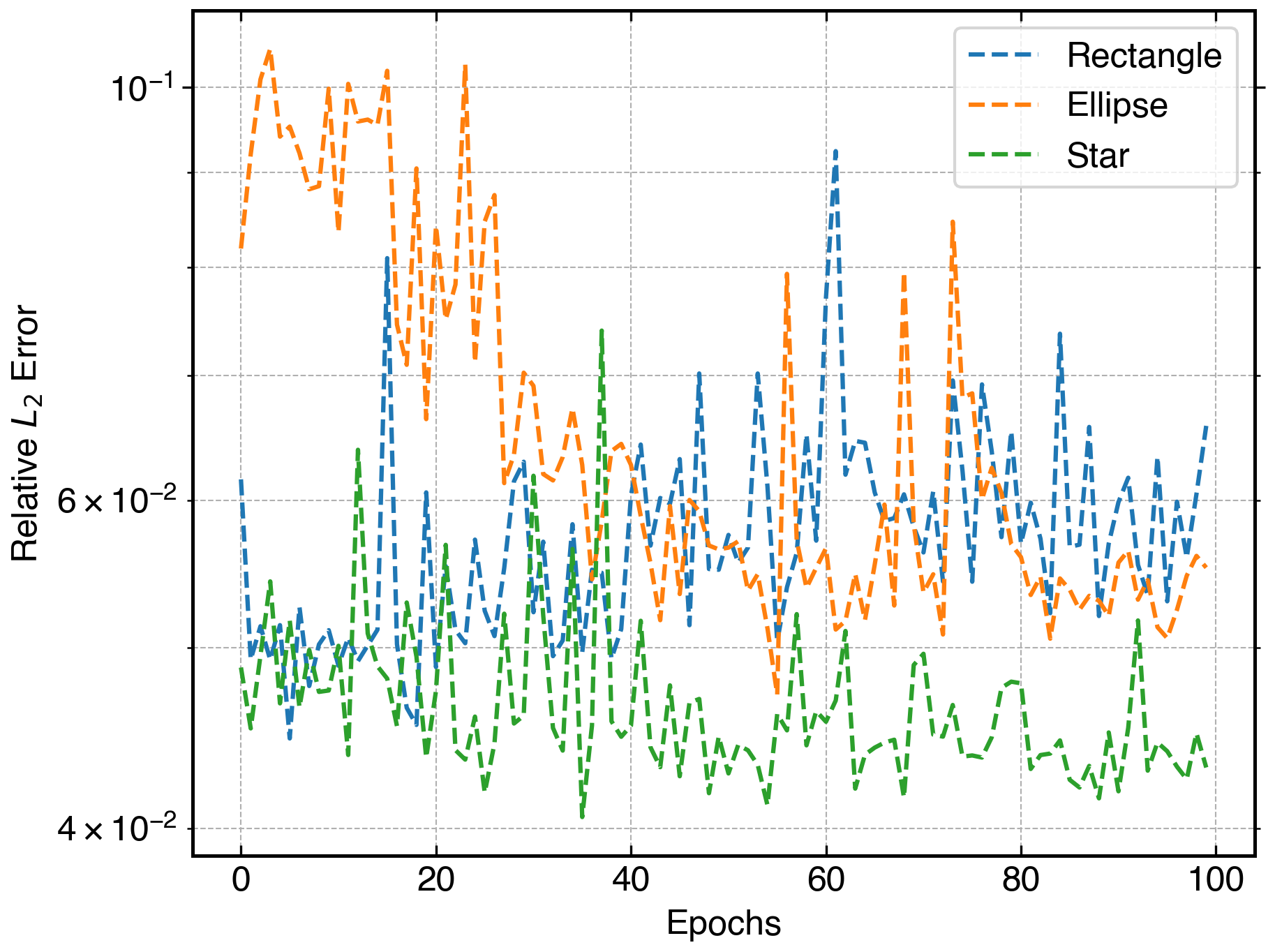}
      \caption{$L_2$ error on test data}
      \label{fig:var_lap_l1_error}
  \end{subfigure}

\caption{Learned Green's function and the $L_2$ error on test data while training using learned Green's function on different domain shapes for the Laplace equation with variable coefficient.}
     \label{fig:var_lap_force}
\end{figure}

\begin{figure}
  \centering
  \begin{subfigure}[b]{0.45\textwidth}
      \centering
      \includegraphics[width=\textwidth]{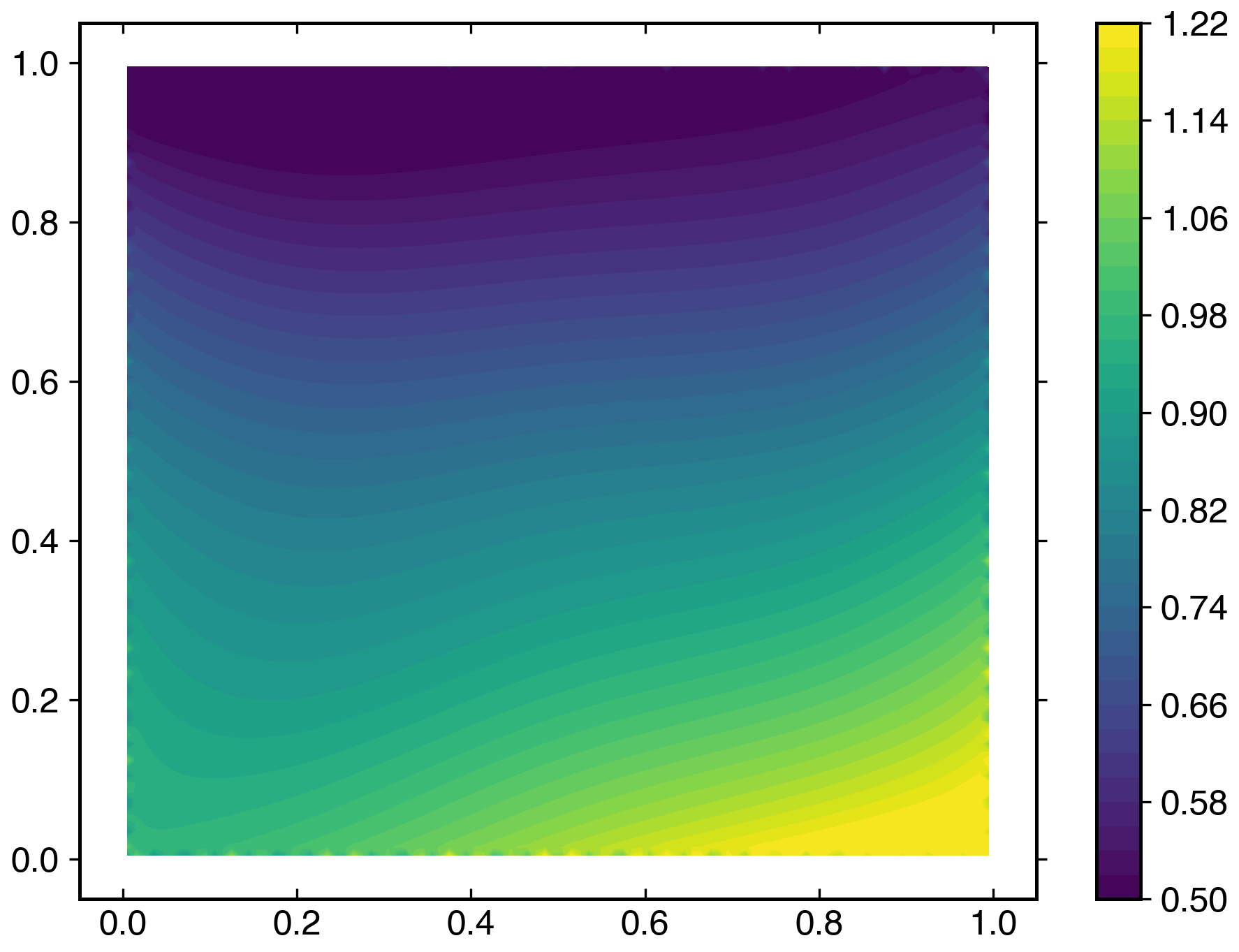}
      \caption{Learned distribution}
      \label{fig:var_lap_gf_rect}
  \end{subfigure}
  \begin{subfigure}[b]{0.45\textwidth}
      \centering
      \includegraphics[width=\textwidth]{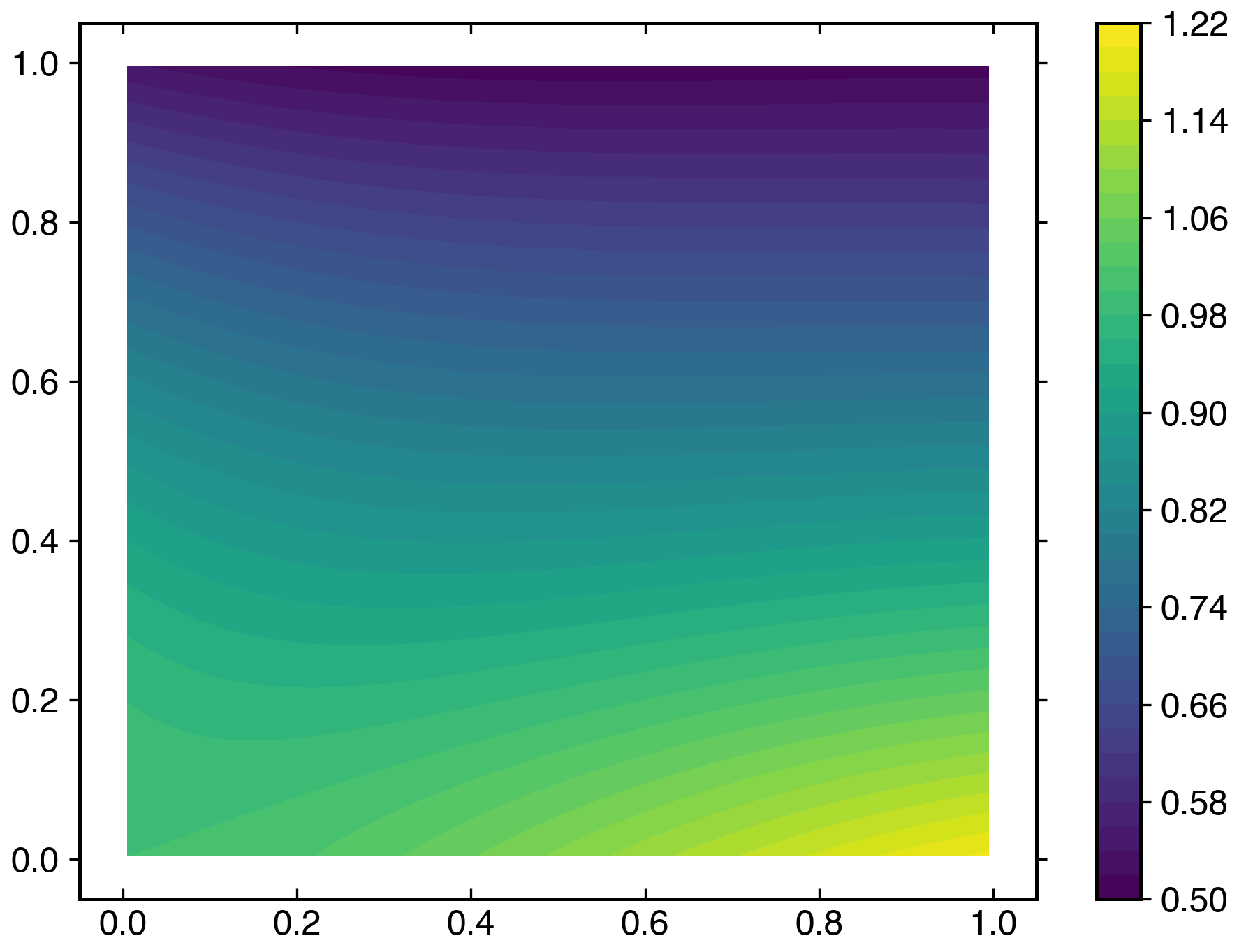}
      \caption{Exact distribution}
      \label{fig:var_lap_gf_err_rect}
  \end{subfigure}
  \begin{subfigure}[b]{0.45\textwidth}
      \centering
      \includegraphics[width=\textwidth]{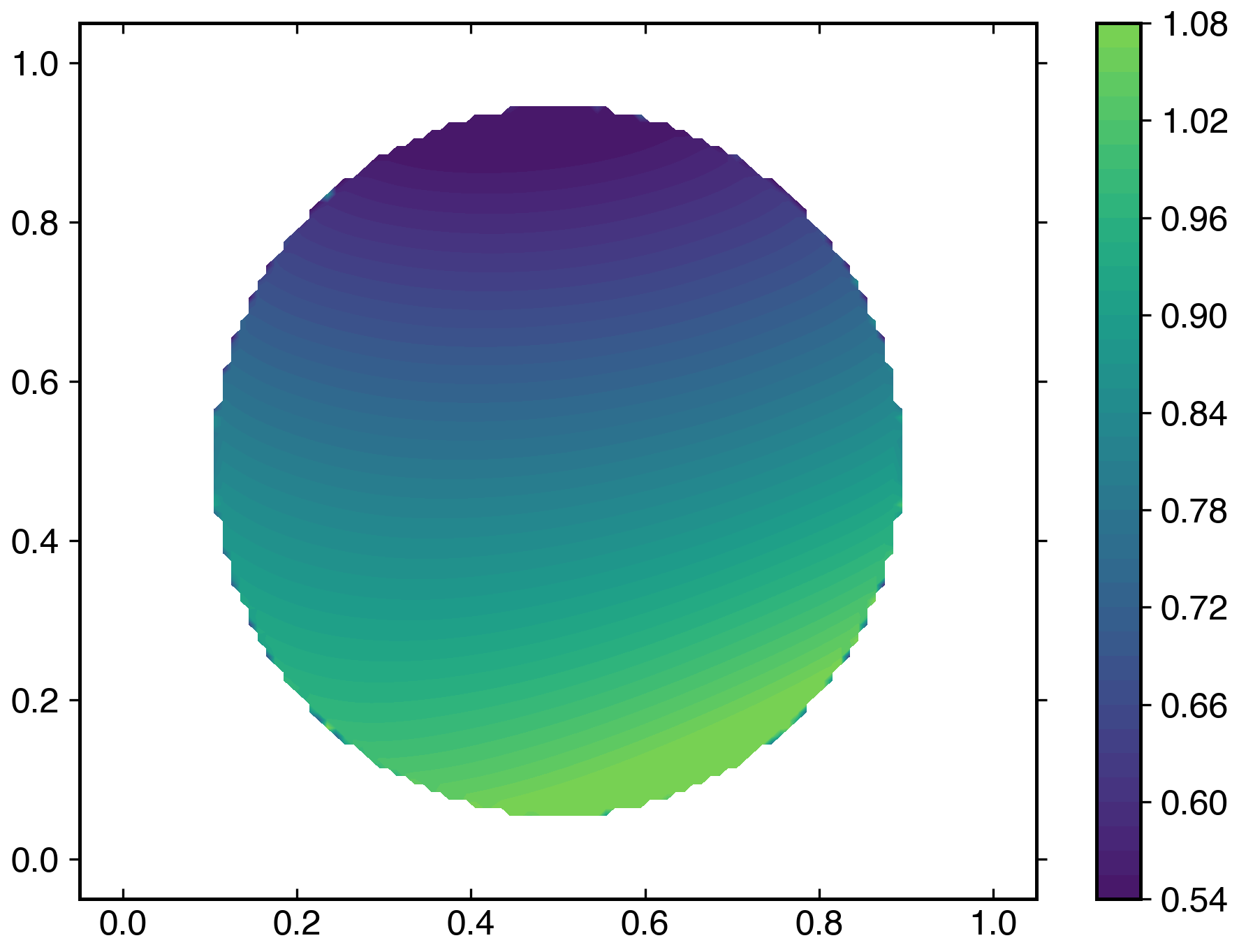}
      \caption{Learned distribution}
      \label{fig:var_lap_gf_ell}
  \end{subfigure}
  \begin{subfigure}[b]{0.45\textwidth}
      \centering
      \includegraphics[width=\textwidth]{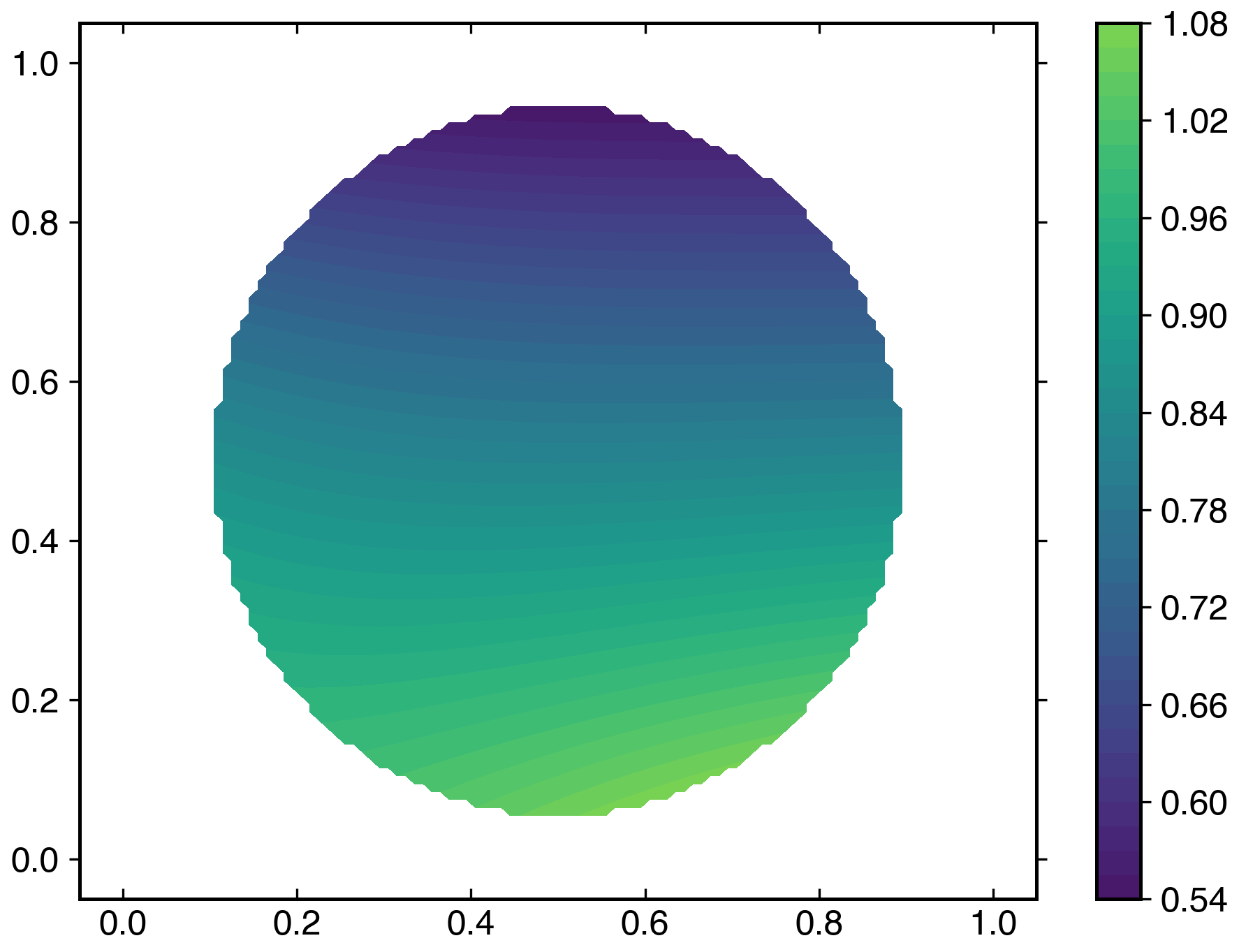}
      \caption{Exact distribution}
      \label{fig:var_lap_gf_err_ell}
  \end{subfigure}
  \begin{subfigure}[b]{0.45\textwidth}
      \centering
      \includegraphics[width=\textwidth]{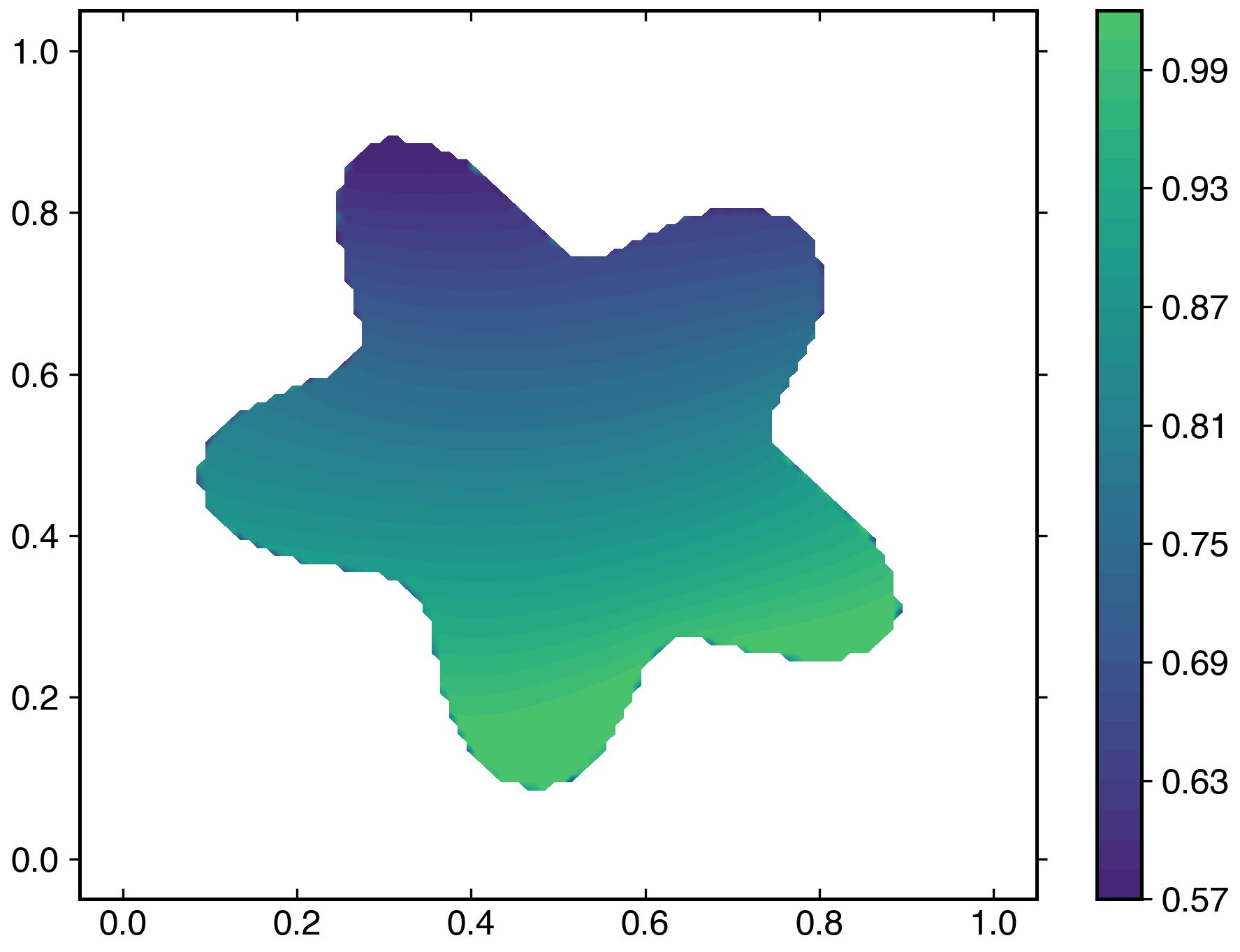}
      \caption{Learned distribution}
      \label{fig:var_lap_gf_star}
  \end{subfigure}
  \begin{subfigure}[b]{0.45\textwidth}
      \centering
      \includegraphics[width=\textwidth]{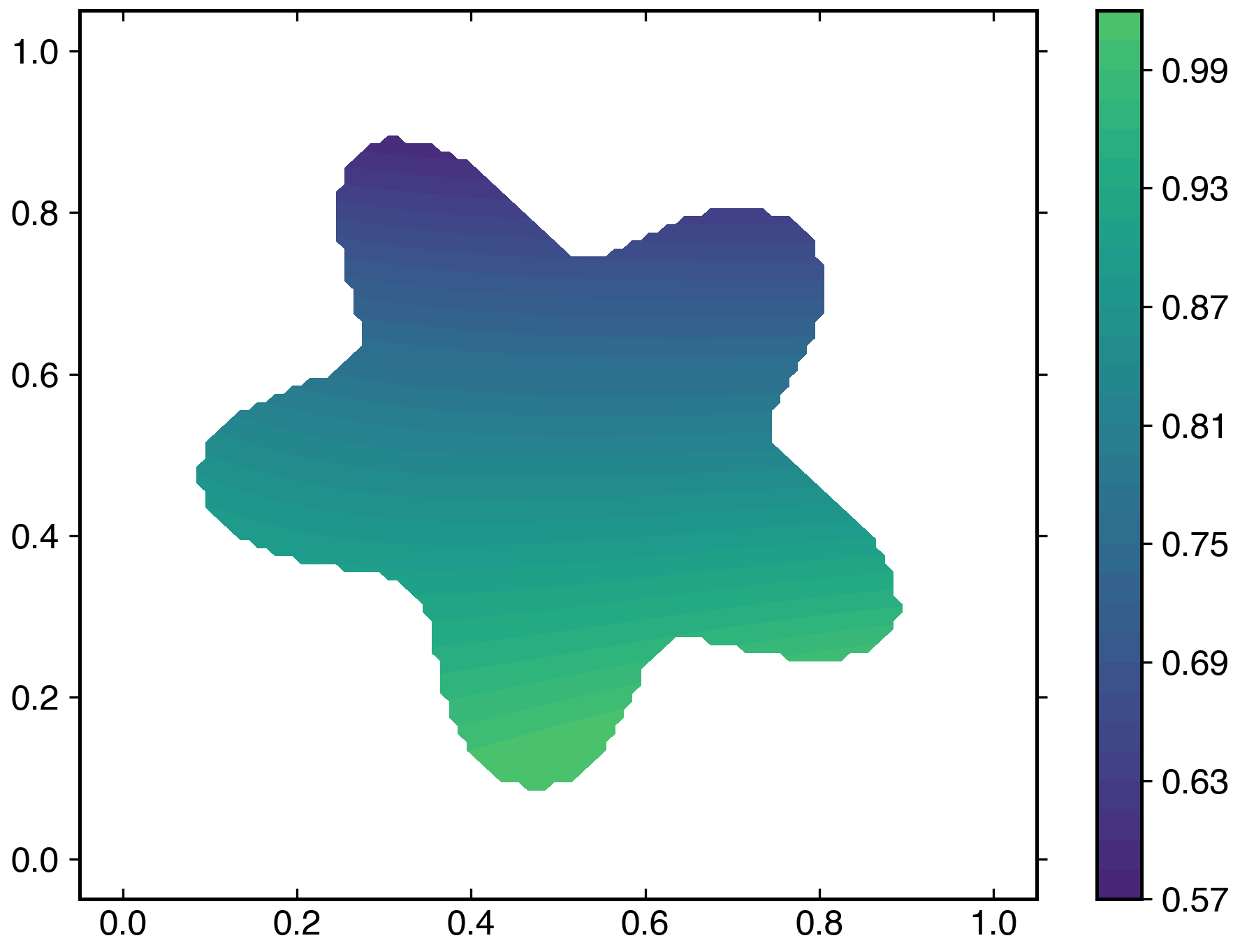}
      \caption{Exact distribution}
      \label{fig:var_lap_gf_err_star}
  \end{subfigure}

\caption{The contour of the learned and expected field using BIN-G for the Laplace equation with variable coefficient.}
     \label{fig:var_lap_force_sols}
\end{figure}

We next consider a Laplace equation with variable coefficient
\begin{equation}
  \nabla \cdot (\sigma(\ten{x}) \nabla u(\ten{x})) = f(\ten{x}),
  \label{eq:var_lap}
\end{equation}
where $\sigma(\ten{x}) = 1.5 + 0.5(\sin(x) + \cos(y))$. We train the Green's function
network using the same methodology and use the learned Green's function to learn density
functions for new domain shapes and test function in \cref{eq:test_func}.
We consider $3$ shapes of the domain viz.\ rectangular, circular, and star
shape. In \cref{fig:var_lap_force}, we plot the learned Green's function and the
convergence of the $L_2$ error on test samples from the domains. The $L_2$
error is within $6\%$. In case of a rectangular domain the errors are
higher due to corner that have a discontinuity in the boundary normals. In
\cref{fig:var_lap_force_sols}, we plot the test function learned and exact
distribution. The learned contour are close to the exact distribution.

\subsection{Helmholtz equation}
\label{sec:helm}

In order to show the capability and ensure the correctness of the proposed
method, we solve the Helmholtz equation for real values. We first learn the
Green's function for the Helmholtz equation
\begin{equation}
  \nabla^2 u(\ten{x}) + k^2 u(\ten{x}) = f(\ten{x}),
  \label{eq:helm_eq}
\end{equation}
where $k$ is the eigenvalue. We choose $k=\{1, 8\}$. We first train the
network to learn the Green's function for the PDE and then employ it to learn the solution
using BIN-G for the test function in \cref{eq:test_func}.

\begin{figure}
  \centering
  \begin{subfigure}[b]{0.45\textwidth}
      \centering
      \includegraphics[width=\textwidth]{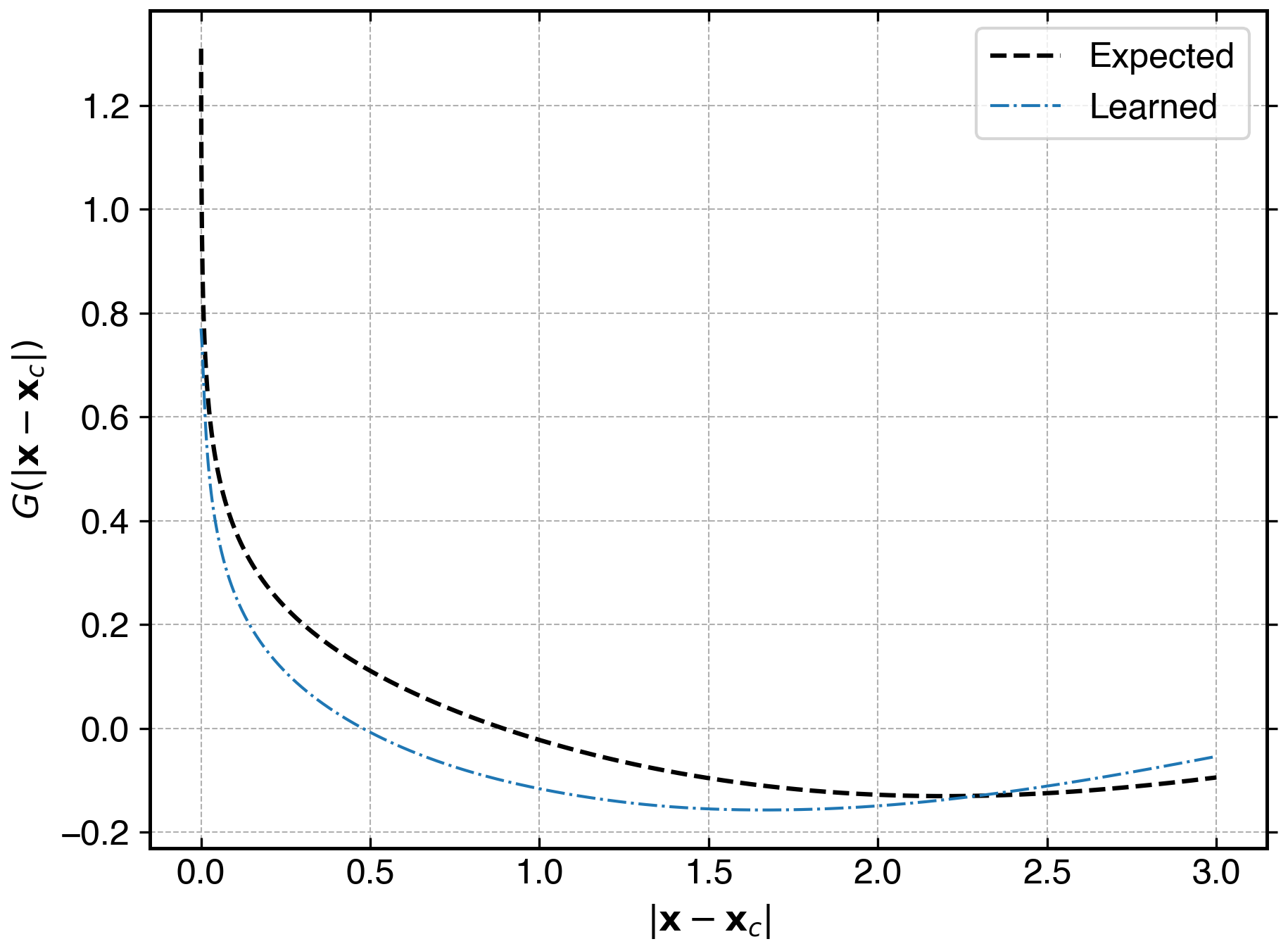}
      \caption{k=1}
      \label{fig:helm_k1}
  \end{subfigure}
  \begin{subfigure}[b]{0.45\textwidth}
      \centering
      \includegraphics[width=\textwidth]{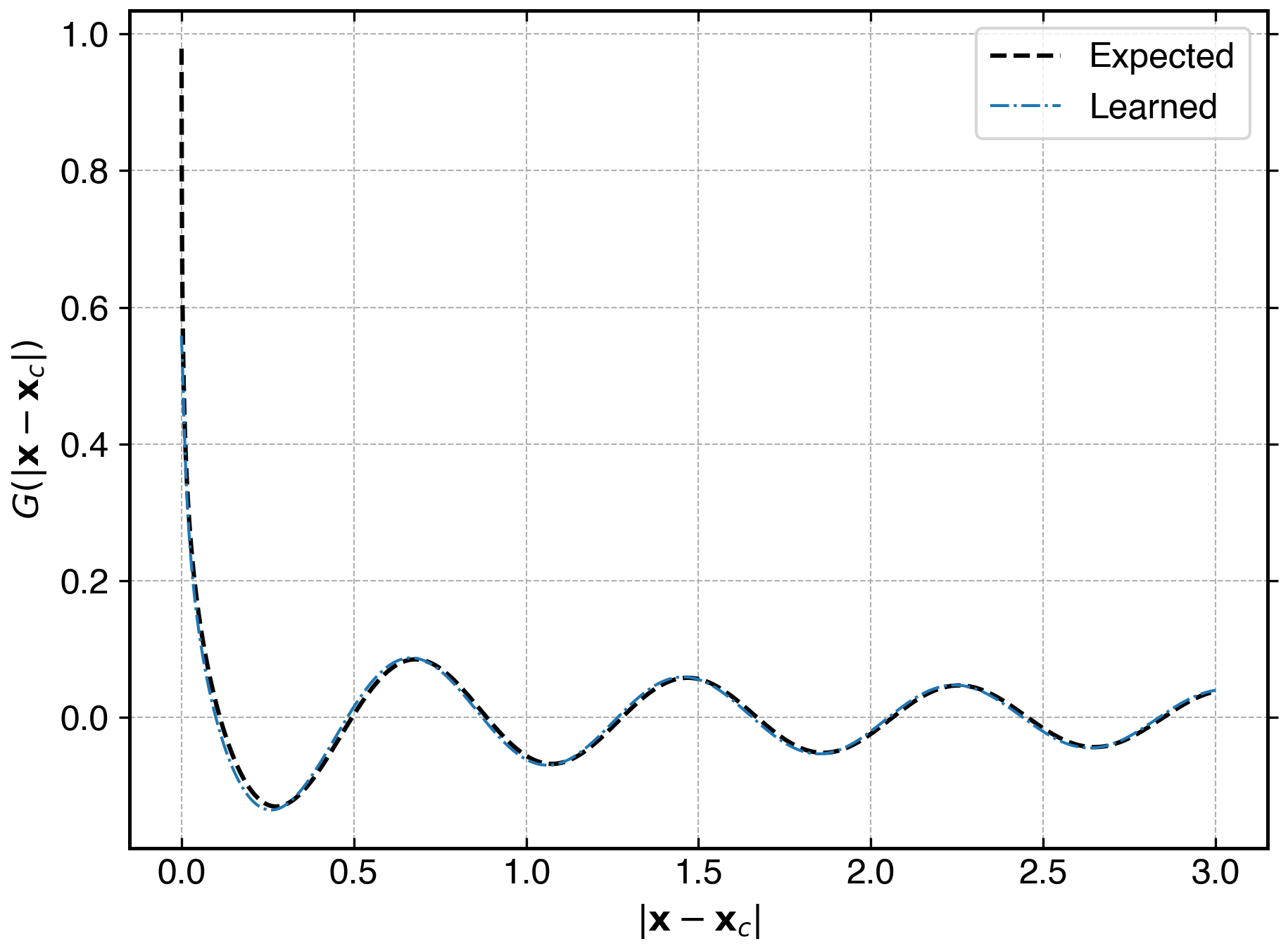}
      \caption{k=8}
      \label{fig:helm_k8}
  \end{subfigure}
\caption{Learned Green's function compared with the analytical Green's function for the Helmholtz equation.}
     \label{fig:helm_gf}
\end{figure}

\begin{figure}
  \centering
  \begin{subfigure}[b]{0.45\textwidth}
      \centering
      \includegraphics[width=\textwidth]{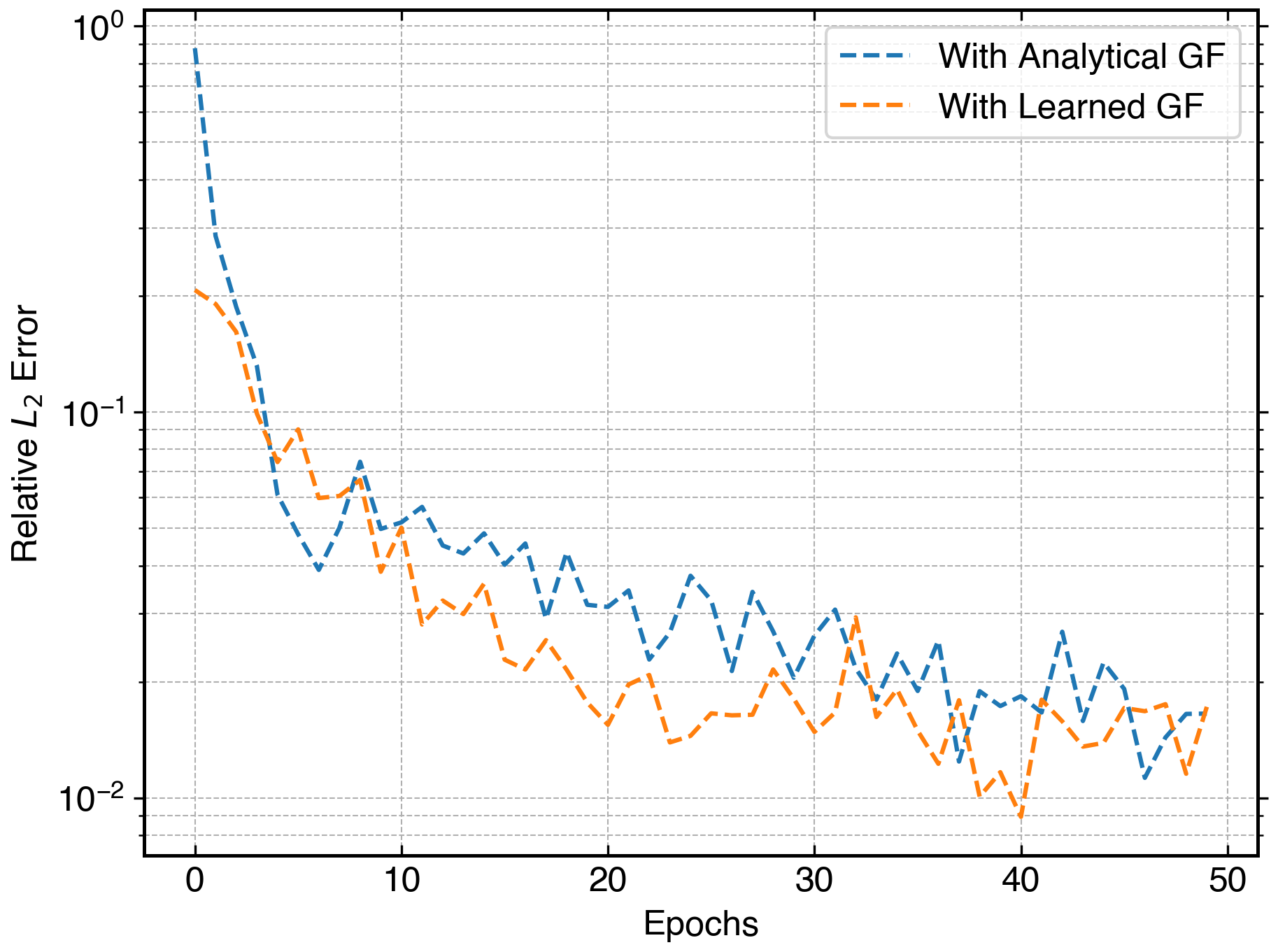}
      \caption{k=1}
      \label{fig:l1_helm_k1}
  \end{subfigure}
  \begin{subfigure}[b]{0.45\textwidth}
      \centering
      \includegraphics[width=\textwidth]{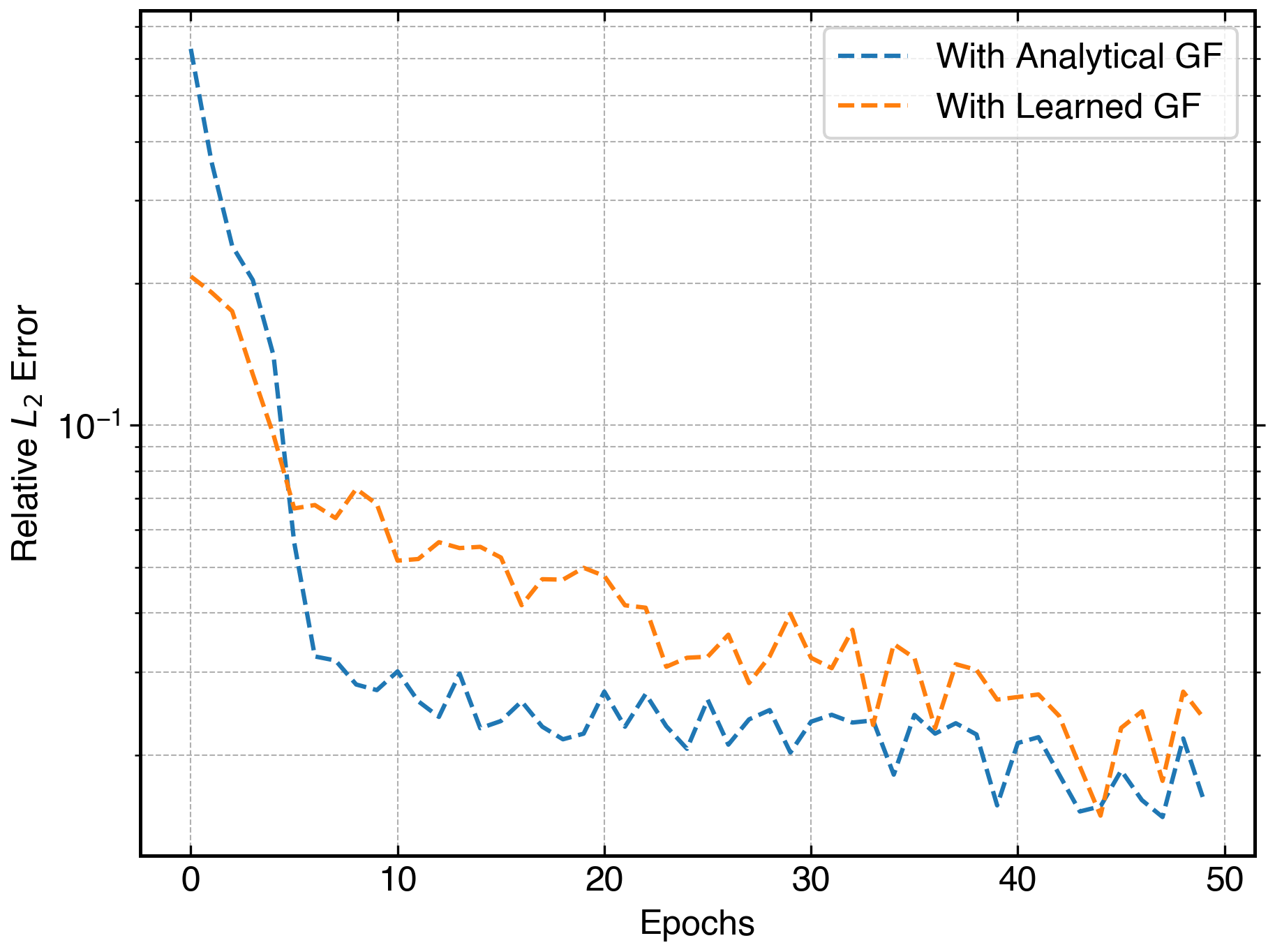}
      \caption{k=8}
      \label{fig:l1_helm_k8}
  \end{subfigure}
\caption{$L_2$ error on test data using BIN-G with learned Green's function
and analytical Green's function for the Helmholtz equation.}
     \label{fig:helm_l1_error}
\end{figure}

In \cref{fig:helm_gf}, we show the comparison of the learned Green's
function with respect to the analytical form. The learned Green's function
shows a close match with the analytical form for $k=8$ compared to $k=1$.
In \cref{fig:helm_l1_error}, we show the comparison of the relative $L_2$
error in the test data while training the density function using an
analytical Green's function and the learned Green's function. The learned Green's function shows a
trend close to the error using an analytical Green's function in the BIN-G. In the case
of $k=1$, the gap between the learned Green's function and the actual Green's function is high.
However, the $L_2$ error while learning density function is close to when
an analytical Green's function is used. This shows that the lower eigensolution are more
likely to converge to wrong solution. We suggest to use a larger domain for
lower eigenvalue PDEs.

\begin{figure}
  \centering
  \begin{subfigure}[b]{0.45\textwidth}
      \centering
      \includegraphics[width=\textwidth]{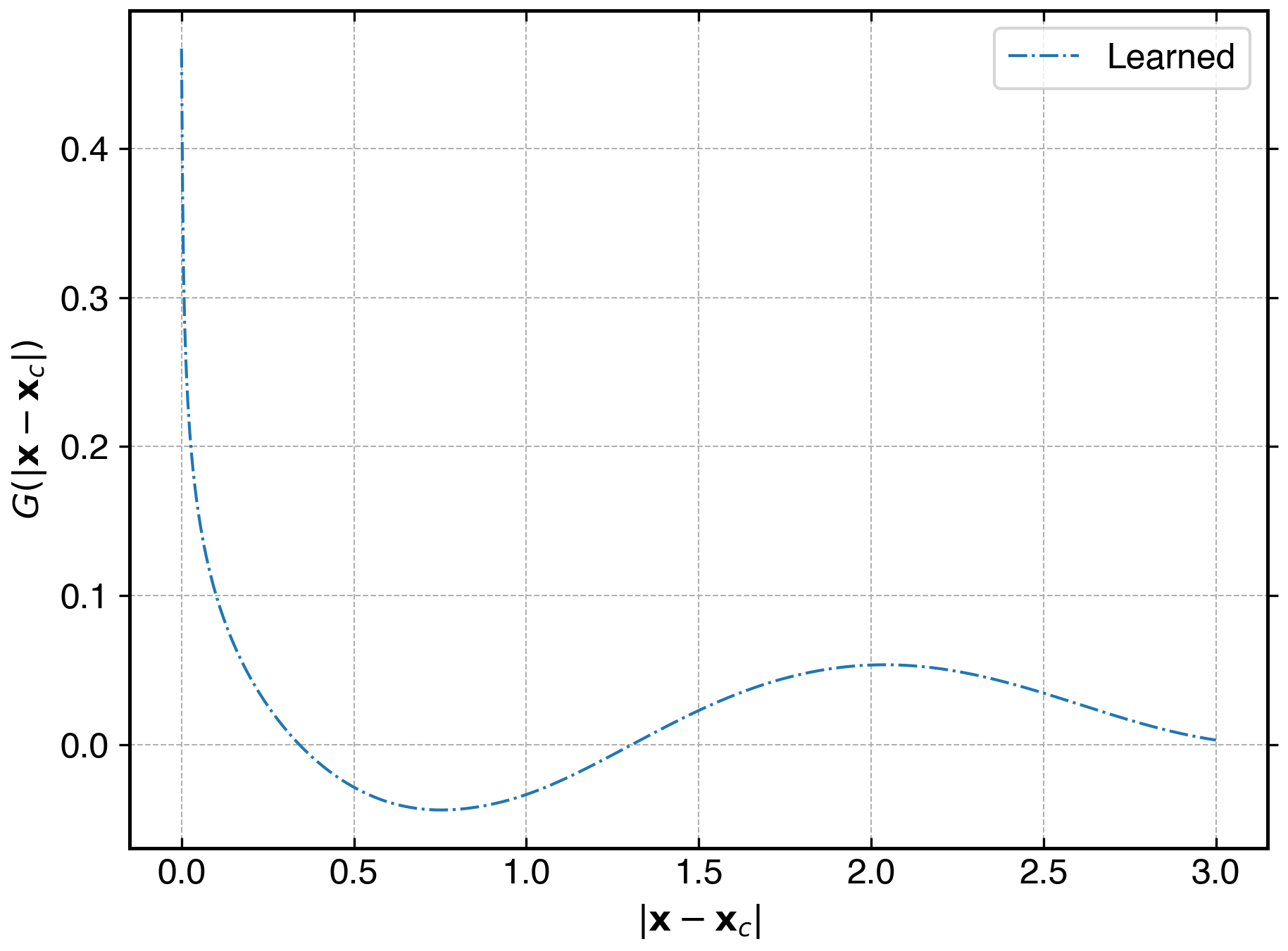}
      \caption{Green's function}
      \label{fig:var_helm_gf}
  \end{subfigure}
  \begin{subfigure}[b]{0.45\textwidth}
      \centering
      \includegraphics[width=\textwidth]{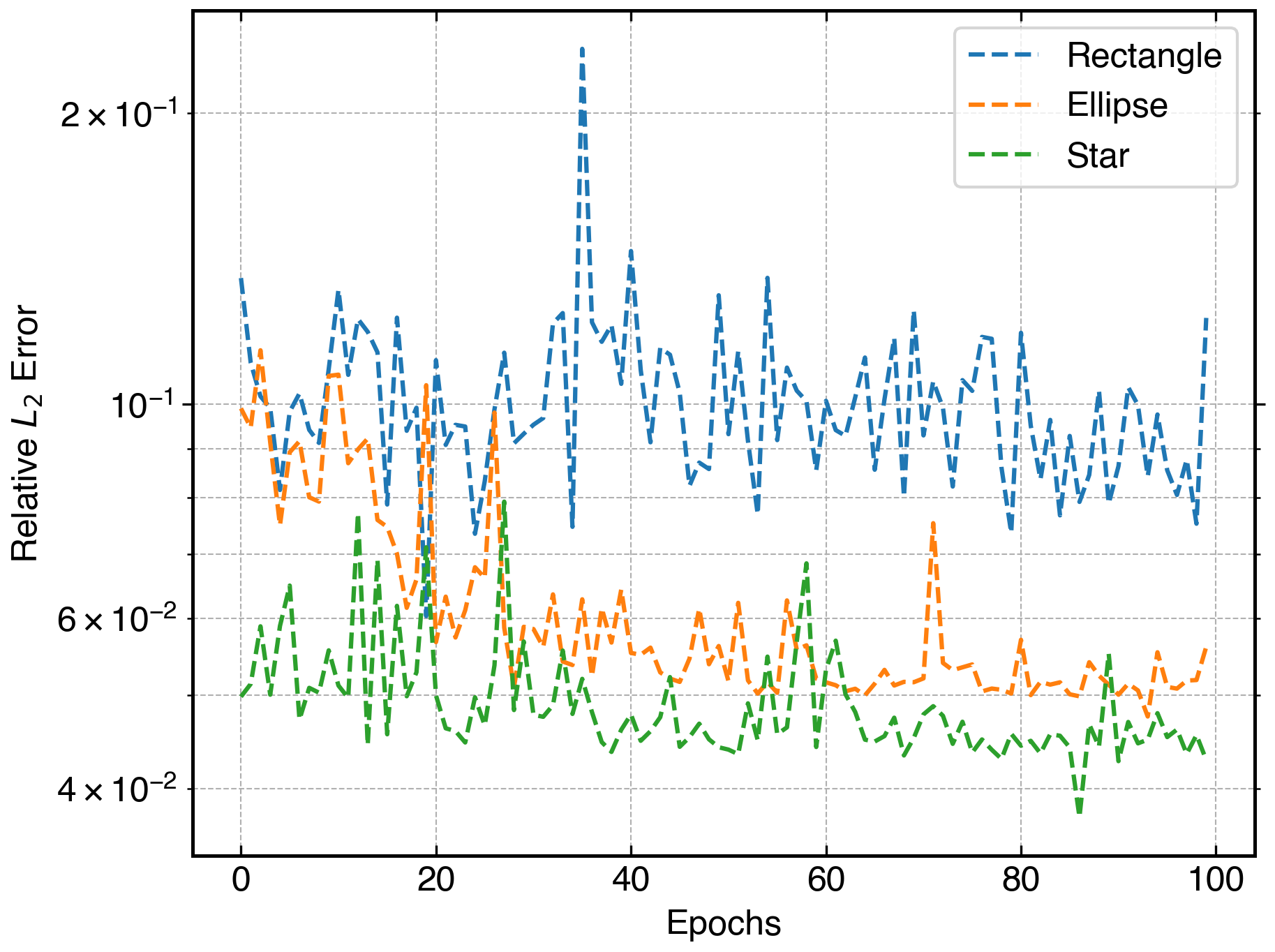}
      \caption{$L_2$ error on test data}
      \label{fig:var_helm_l1_error}
  \end{subfigure}
\caption{Learned Green's function and the $L_2$ error on test data while training using learned Green's function on different domain shapes for the Helmholtz equation with variable coefficient and $k=4$.}
     \label{fig:var_helm_force}
\end{figure}

\begin{figure}
  \centering
  \begin{subfigure}[b]{0.45\textwidth}
      \centering
      \includegraphics[width=\textwidth]{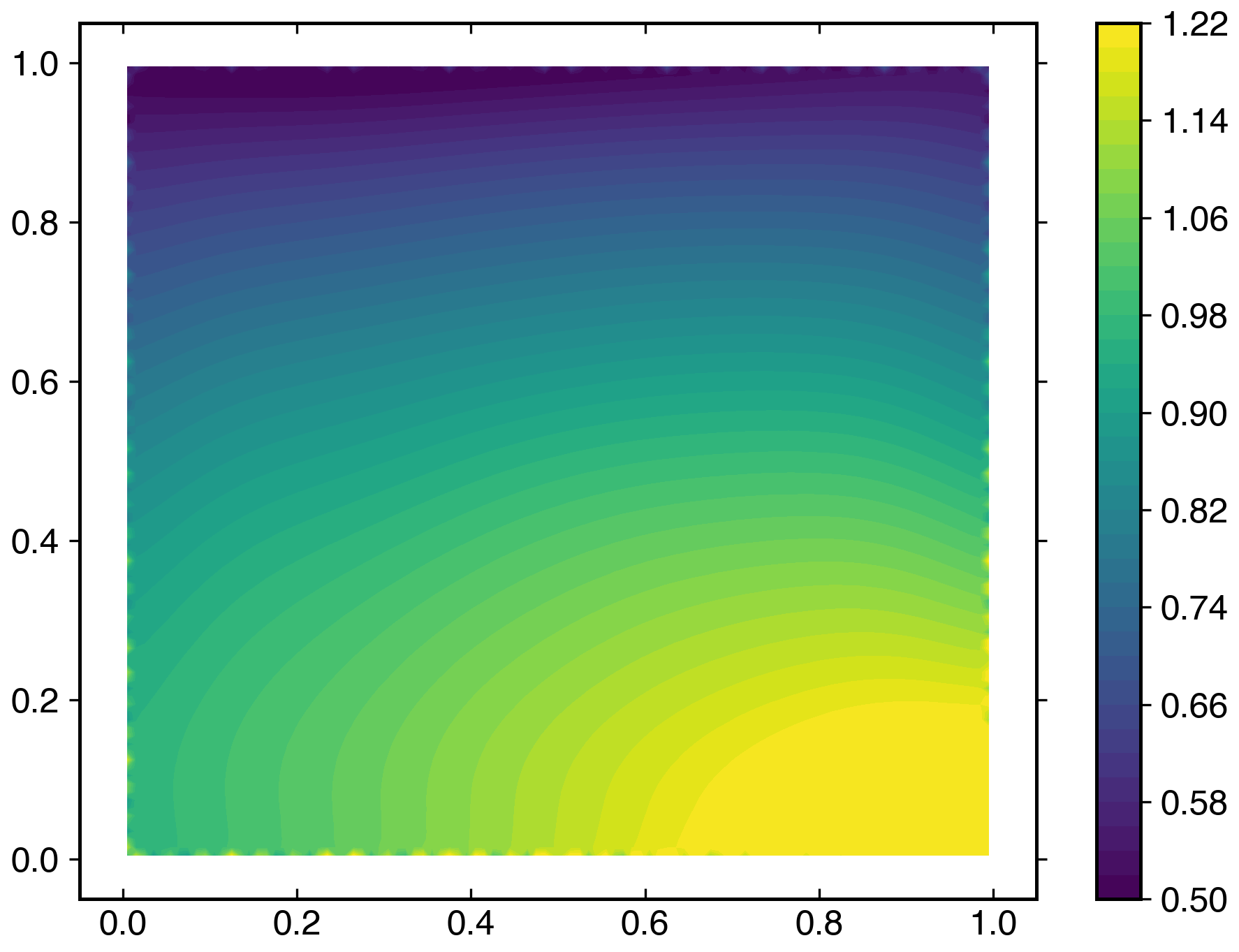}
      \caption{Learned distribution}
      \label{fig:var_helm_gf_rect}
  \end{subfigure}
  \begin{subfigure}[b]{0.45\textwidth}
      \centering
      \includegraphics[width=\textwidth]{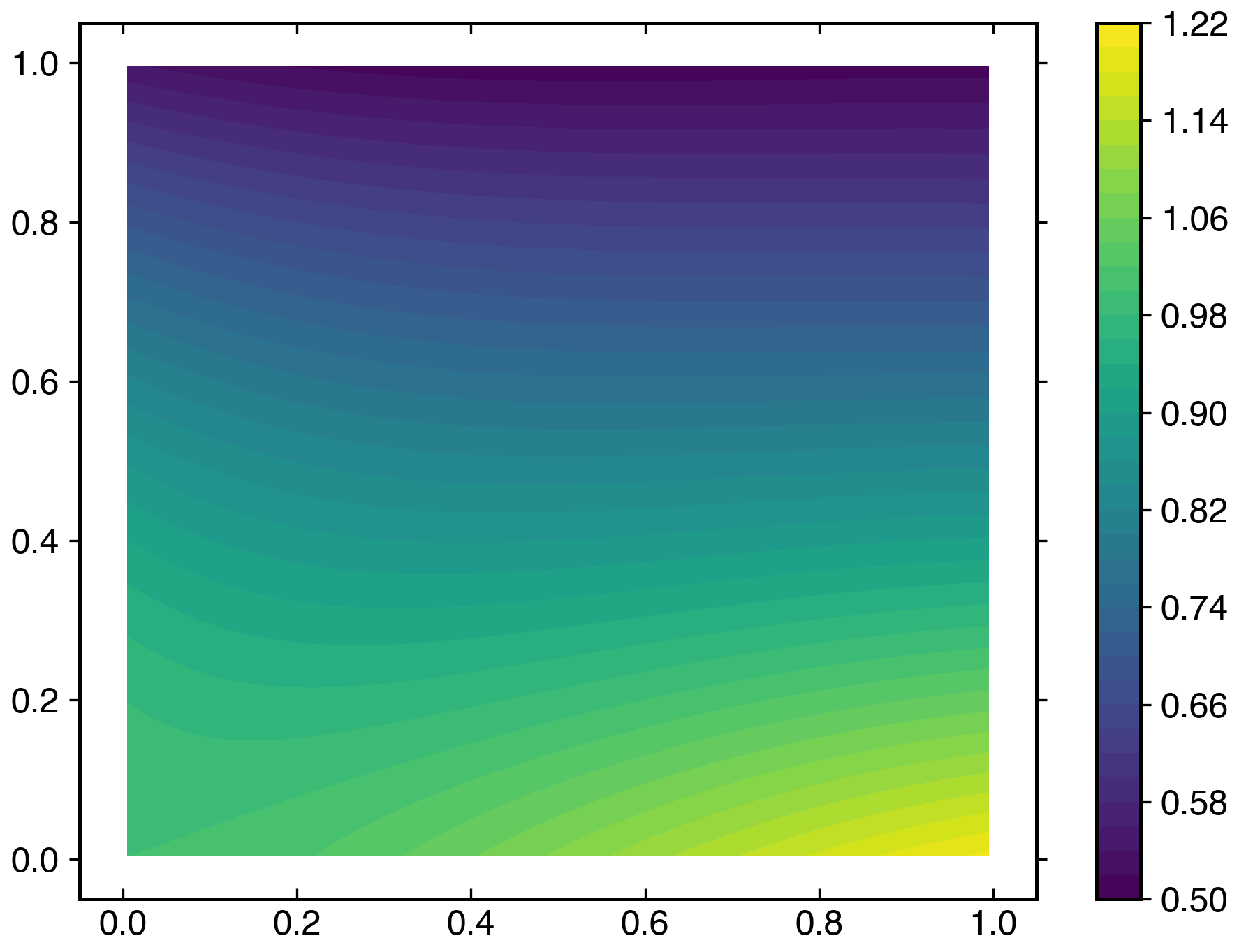}
      \caption{Exact distribution}
      \label{fig:var_helm_gf_err_rect}
  \end{subfigure}
  \begin{subfigure}[b]{0.45\textwidth}
      \centering
      \includegraphics[width=\textwidth]{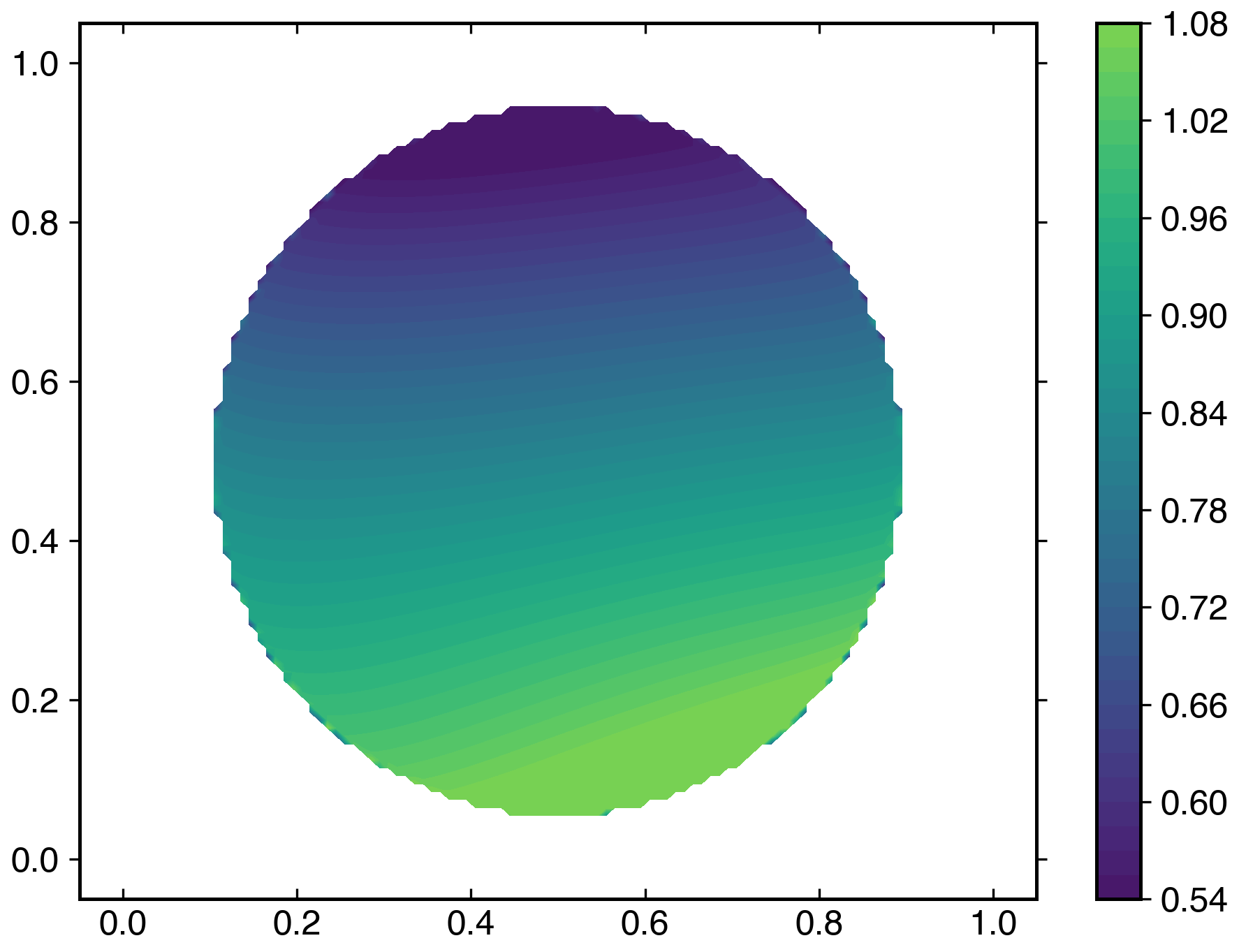}
      \caption{Learned distribution}
      \label{fig:var_helm_gf_ell}
  \end{subfigure}
  \begin{subfigure}[b]{0.45\textwidth}
      \centering
      \includegraphics[width=\textwidth]{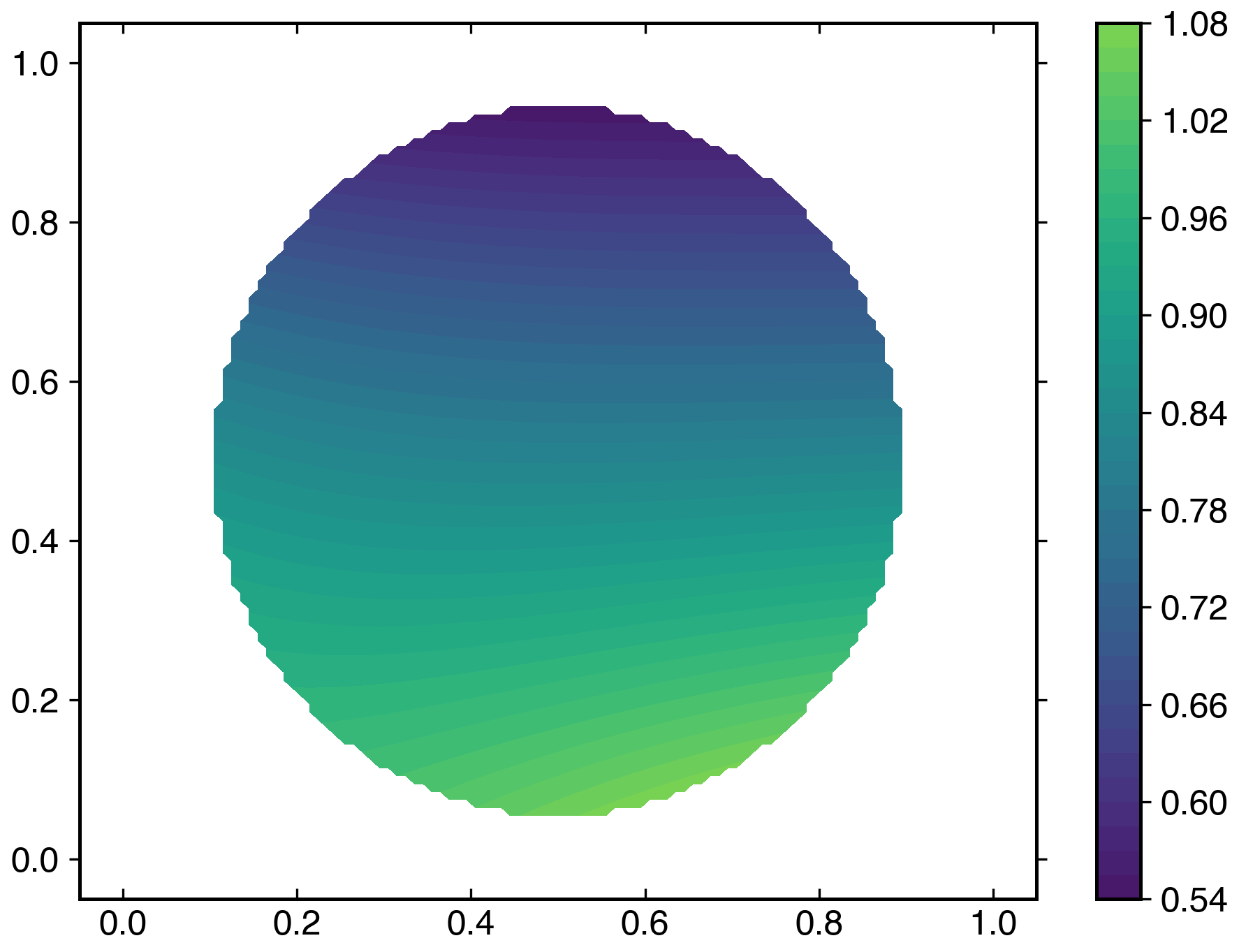}
      \caption{Exact distribution}
      \label{fig:var_helm_gf_err_ell}
  \end{subfigure}
  \begin{subfigure}[b]{0.45\textwidth}
      \centering
      \includegraphics[width=\textwidth]{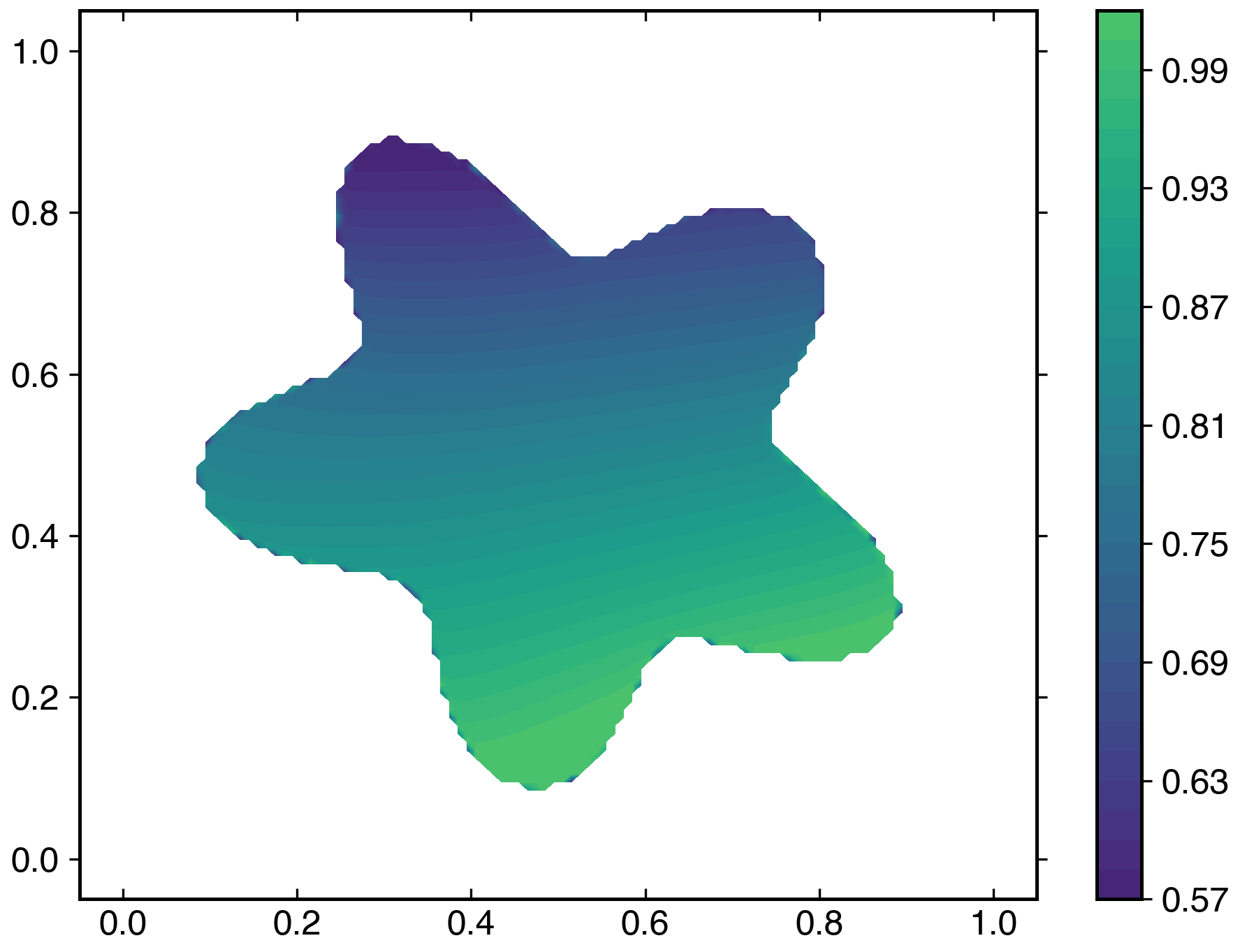}
      \caption{Learned distribution}
      \label{fig:var_helm_gf_star}
  \end{subfigure}
  \begin{subfigure}[b]{0.45\textwidth}
      \centering
      \includegraphics[width=\textwidth]{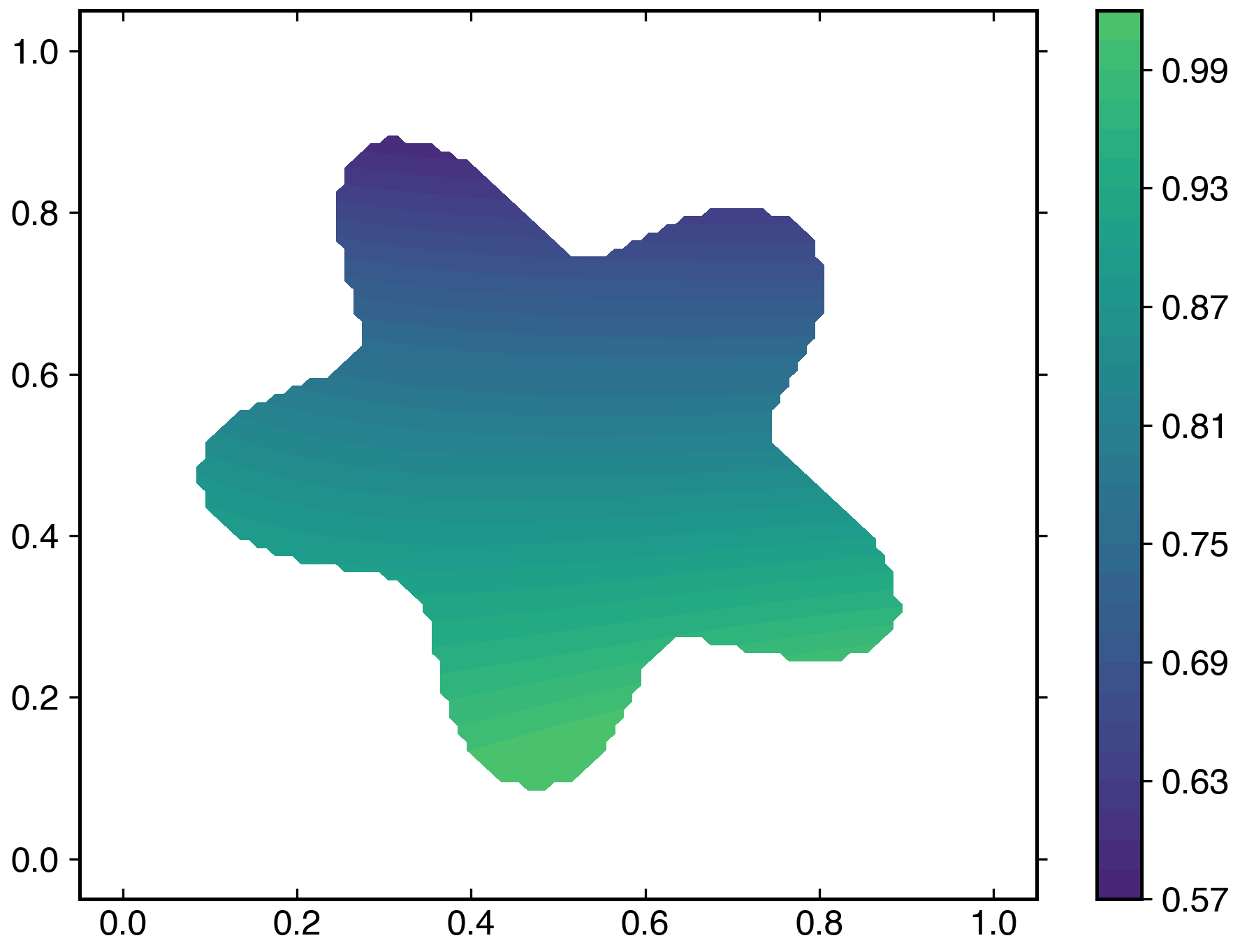}
      \caption{Exact distribution}
      \label{fig:var_helm_gf_err_star}
  \end{subfigure}
\caption{The contour of the learned field using BIN-G and the error distribution for the Helmholtz equation with variable coefficient and $k=4$.}
     \label{fig:var_hel_force_sols}
\end{figure}

We also consider a variable coefficient Helmholtz equation with constant
eigenvalue given by
\begin{equation}
  \nabla \cdot (\sigma(\ten{x}) \nabla u(\ten{x})) + k^2 u(\ten{x}) = f(\ten{x}),
  \label{eq:var_helm}
\end{equation}
where we set $k=4$. We learn the Green's function for the above PDE using
the present method and employ it to learn the solution for the test function
as done in previous test cases. However, we consider different domain
shapes viz.\ rectangular, ellipse, and star. In \cref{fig:var_helm_force},
we plot the learned Green's function and the relative $L_2$ error in the
test data while training for different domains. As observed in the case of
the Laplace equation, the errors in the case of the rectangular domain is high
due to corners. The error in the case of Elliptical and star shaped domain
are within $5\%$. These errors can be improved by using a better
quadrature, larger domain, and more number of source-test function pairs.
In \cref{fig:var_hel_force_sols}, we plot the learned solution and exact
distribution for different domains. The higher error in the case of
the rectangular domain is visible by higher values in the distribution.
However, in the case of the other two domains the contours are close to the
expected distribution.

\subsection{PDE with variable coefficient}
\label{sec:var_coeff}

In this section, we solve a variable coefficient elliptic PDE
\cite{caoKernelfreeBoundaryIntegral2022} given by
\begin{equation}
  \nabla \cdot (\sigma(\ten{x}) \nabla u(\ten{x})) - \kappa(\ten{x}) u(\ten{x}) = f(\ten{x}),
  \label{eq:var_coeff}
\end{equation}
\begin{sloppypar}
where $\sigma(x, y) = 1.5 + 0.5 (\sin(x) + \cos(y))$, and $\kappa(x, y) =
20 + \exp(1.5 x + 1.8 y)$. We note that an analytical form of Green's
function for variable coefficient PDEs are not known. We learn the Green's
function for the PDE in \cref{eq:var_coeff} as done in previous test case.
We use the learned Green's function to solve PDE for an arbitrary test
function on different domain shapes.

\end{sloppypar}

\begin{figure}
  \centering
  \begin{subfigure}[b]{0.45\textwidth}
      \centering
      \includegraphics[width=\textwidth]{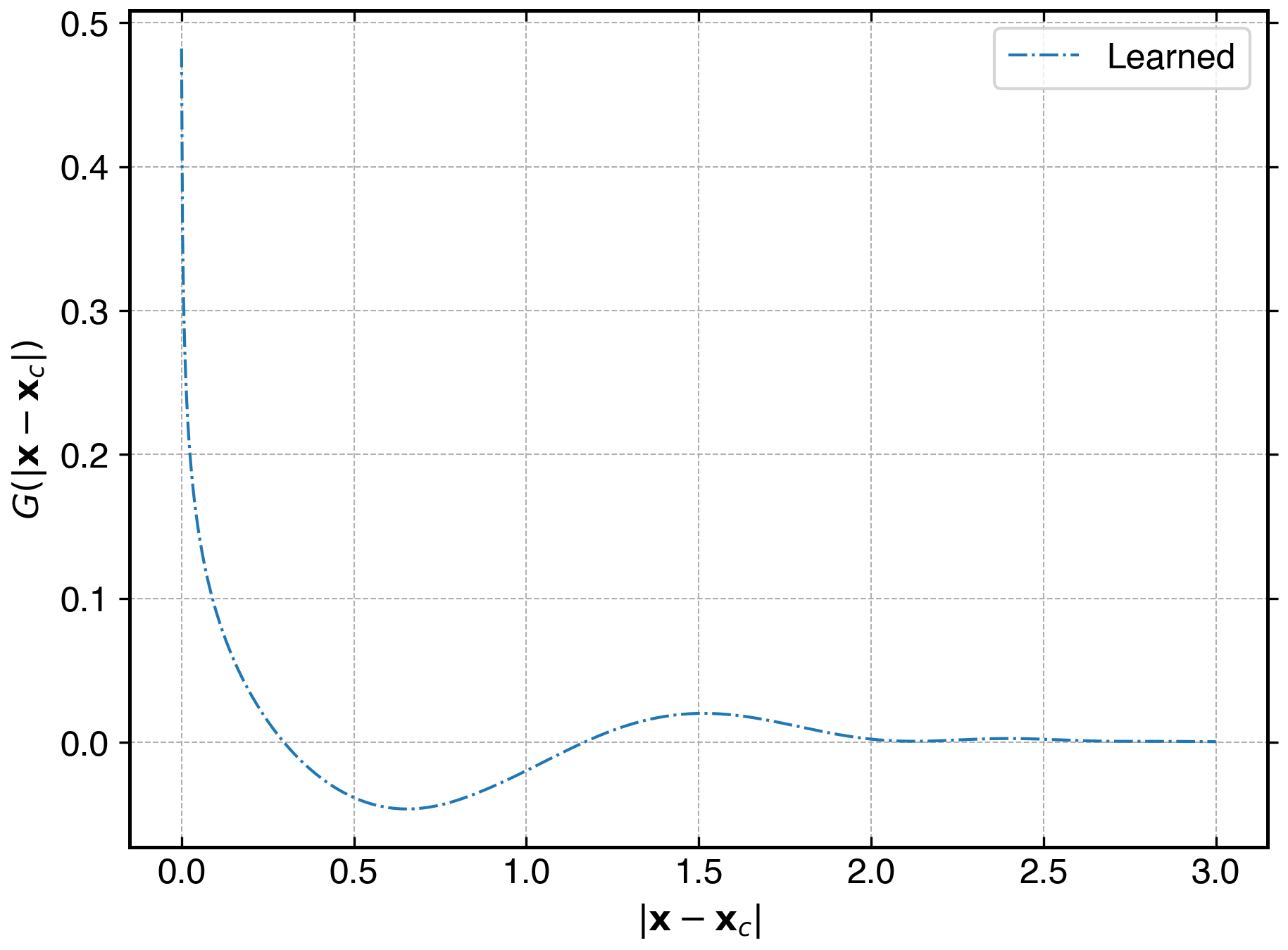}
      \caption{Green's function}
      \label{fig:var_coeff_gf}
  \end{subfigure}
  \begin{subfigure}[b]{0.45\textwidth}
      \centering
      \includegraphics[width=\textwidth]{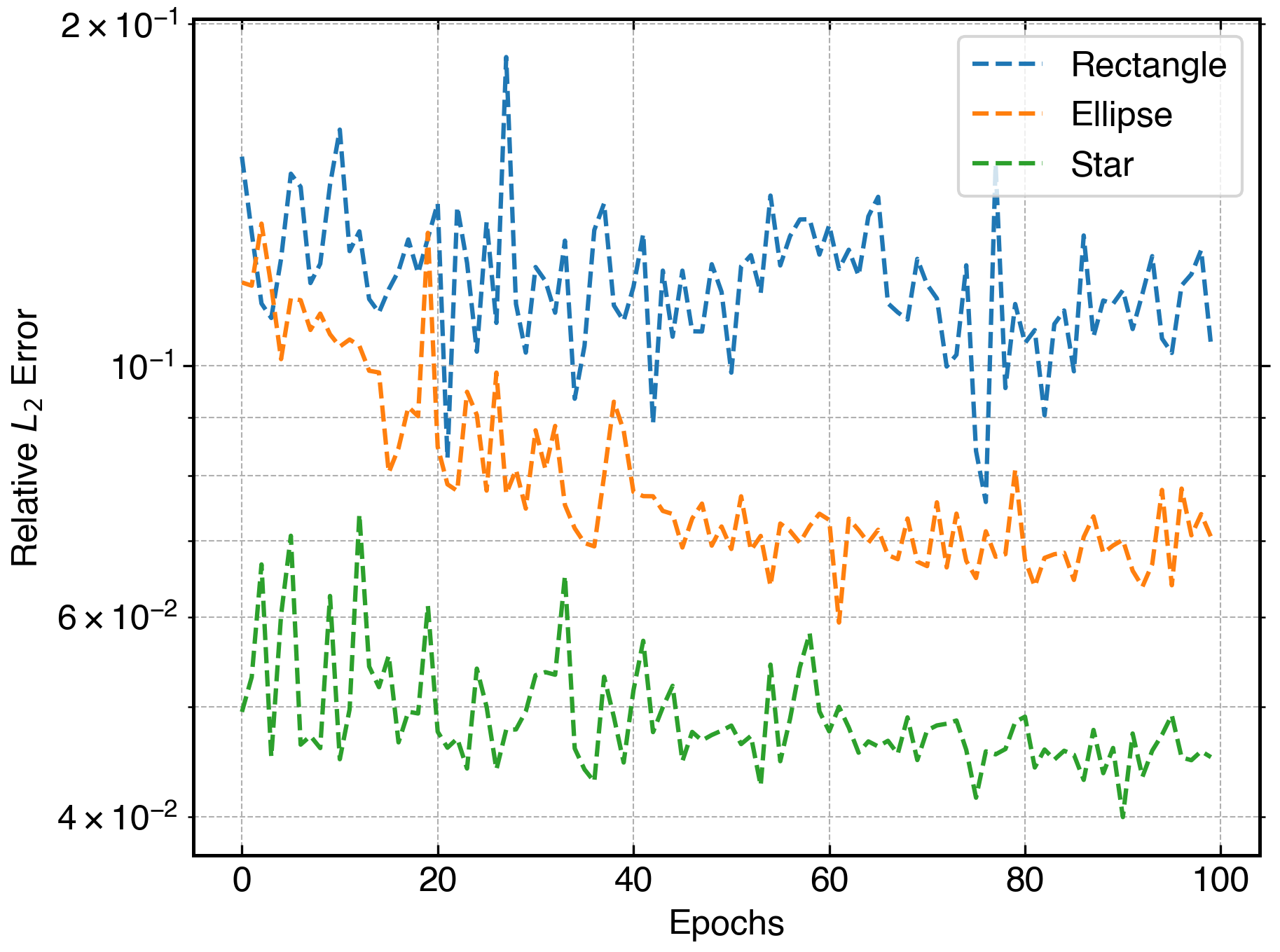}
      \caption{$L_1$ error on test data}
      \label{fig:var_coeff_l1_error}
  \end{subfigure}
\caption{Learned Green's function and the $L_2$ error on test data while training using learned Green's function on different domain shapes for the elliptical equation with variable $\sigma$ and $\kappa$.}
     \label{fig:var_coeff}
\end{figure}

\begin{figure}
  \centering
  \begin{subfigure}[b]{0.45\textwidth}
      \centering
      \includegraphics[width=\textwidth]{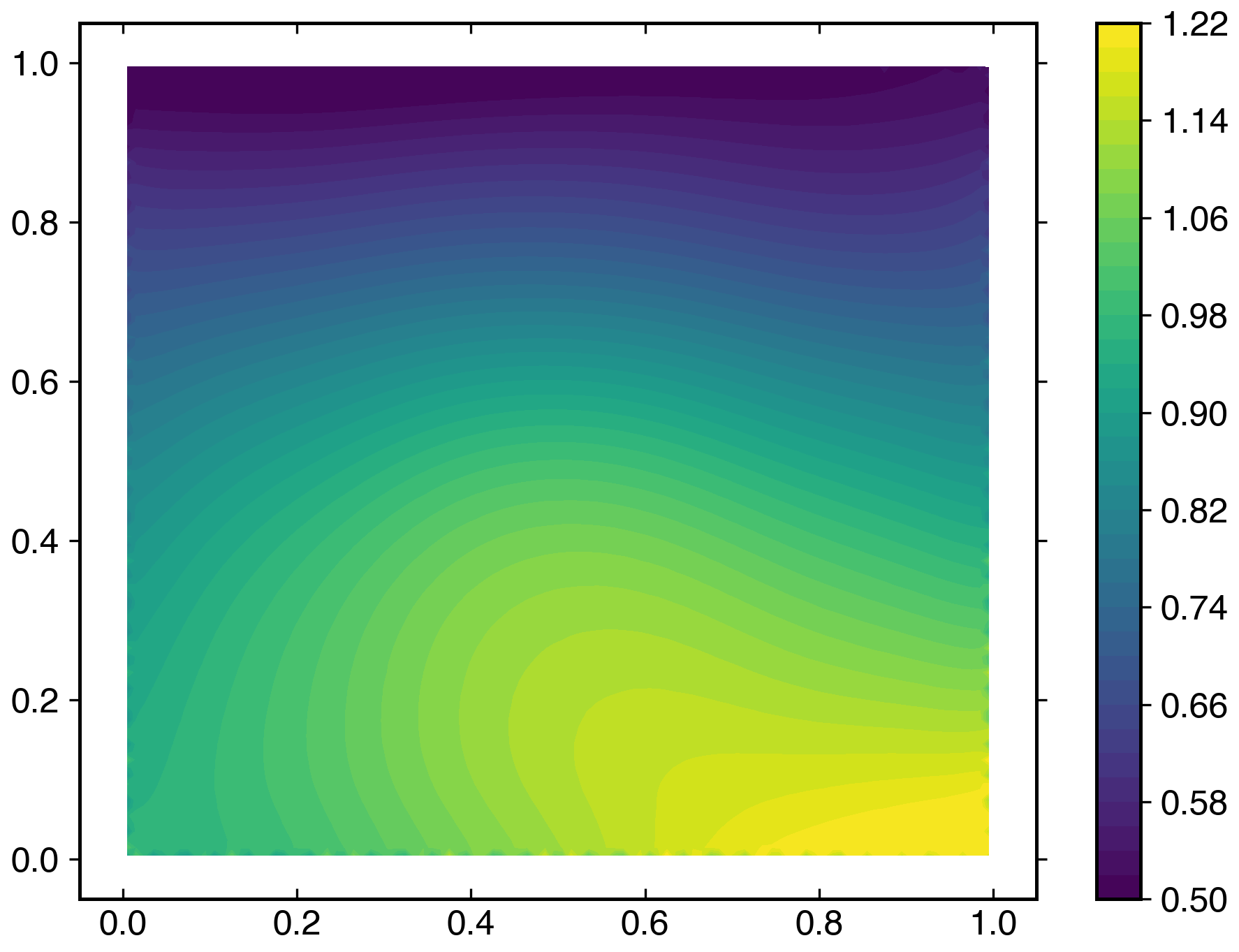}
      \caption{Learned distribution}
      \label{fig:var_coeff_gf_rect}
  \end{subfigure}
  \begin{subfigure}[b]{0.45\textwidth}
      \centering
      \includegraphics[width=\textwidth]{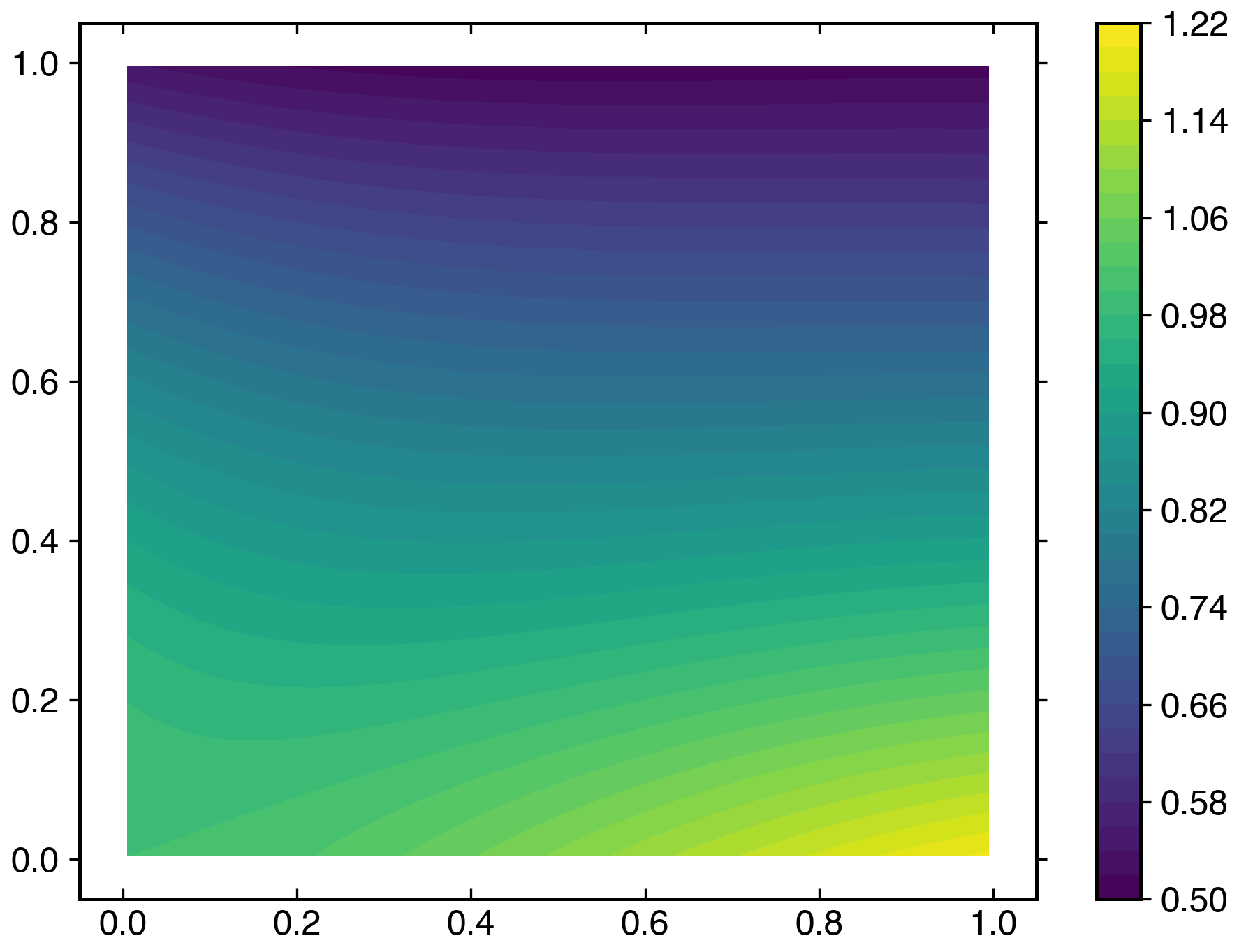}
      \caption{Exact distribution}
      \label{fig:var_coeff_gf_err_rect}
  \end{subfigure}
  \begin{subfigure}[b]{0.45\textwidth}
      \centering
      \includegraphics[width=\textwidth]{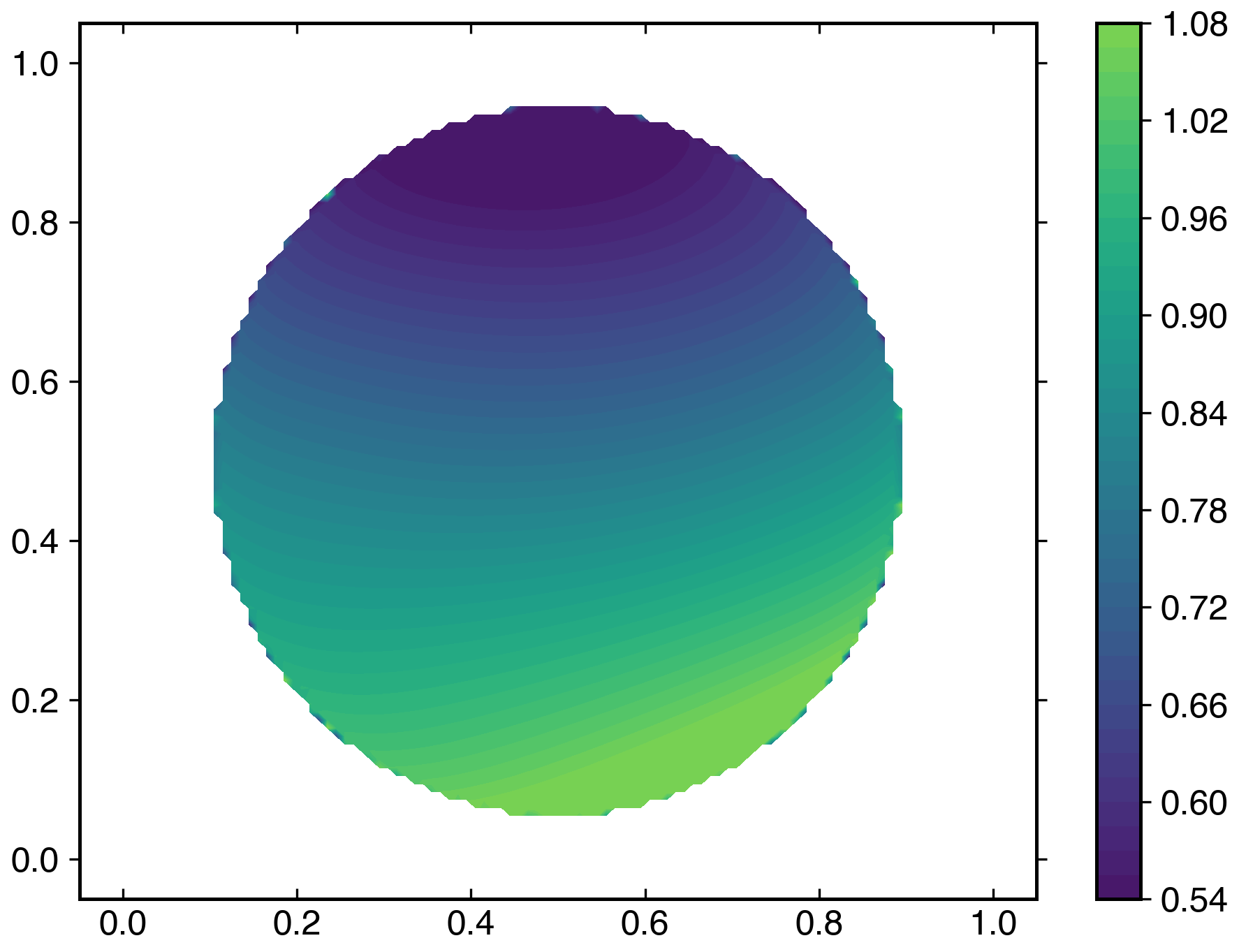}
      \caption{Learned distribution}
      \label{fig:var_coeff_gf_ell}
  \end{subfigure}
  \begin{subfigure}[b]{0.45\textwidth}
      \centering
      \includegraphics[width=\textwidth]{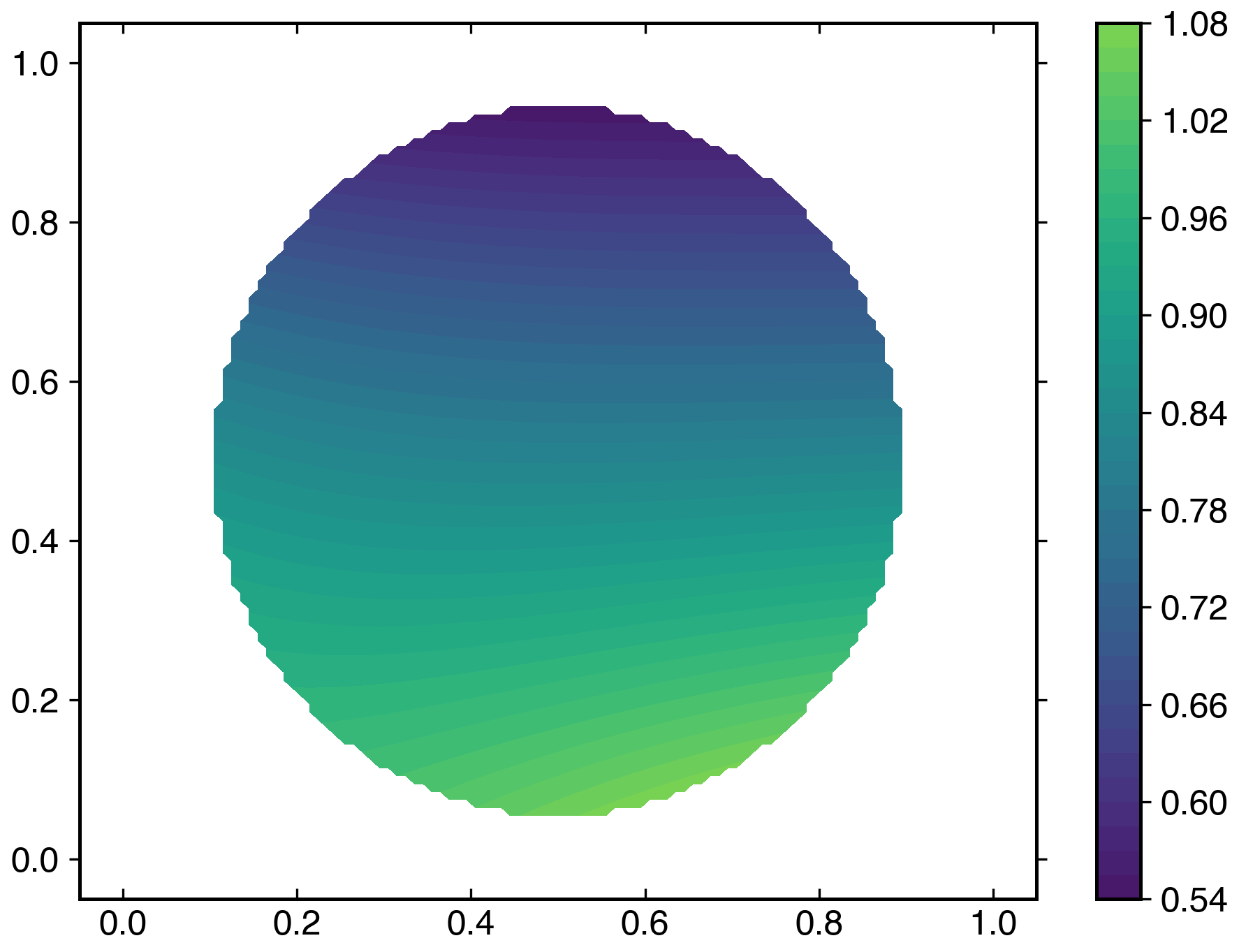}
      \caption{Exact distribution}
      \label{fig:var_coeff_gf_err_ell}
  \end{subfigure}
  \begin{subfigure}[b]{0.45\textwidth}
      \centering
      \includegraphics[width=\textwidth]{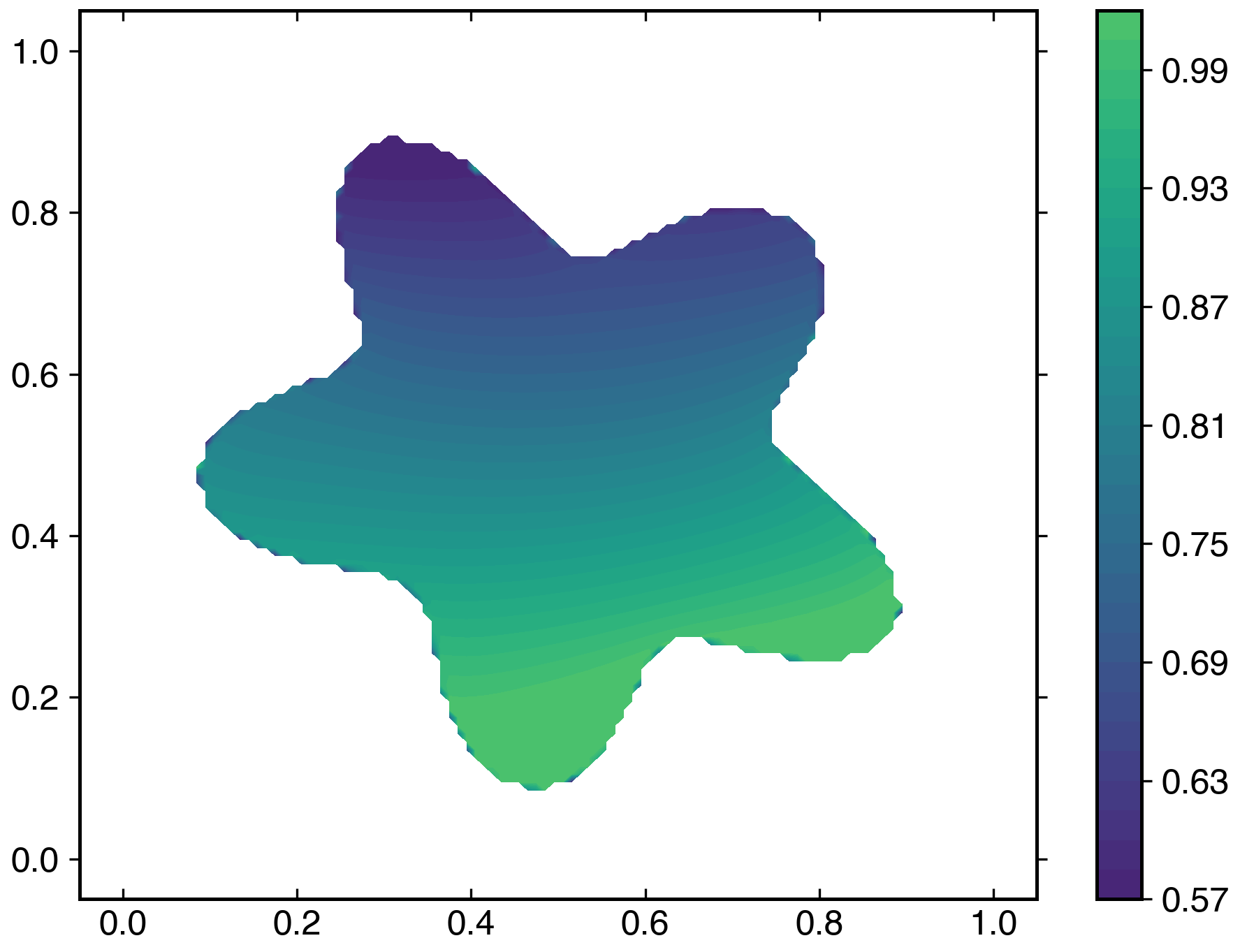}
      \caption{Learned distribution}
      \label{fig:var_coeff_gf_star}
  \end{subfigure}
  \begin{subfigure}[b]{0.45\textwidth}
      \centering
      \includegraphics[width=\textwidth]{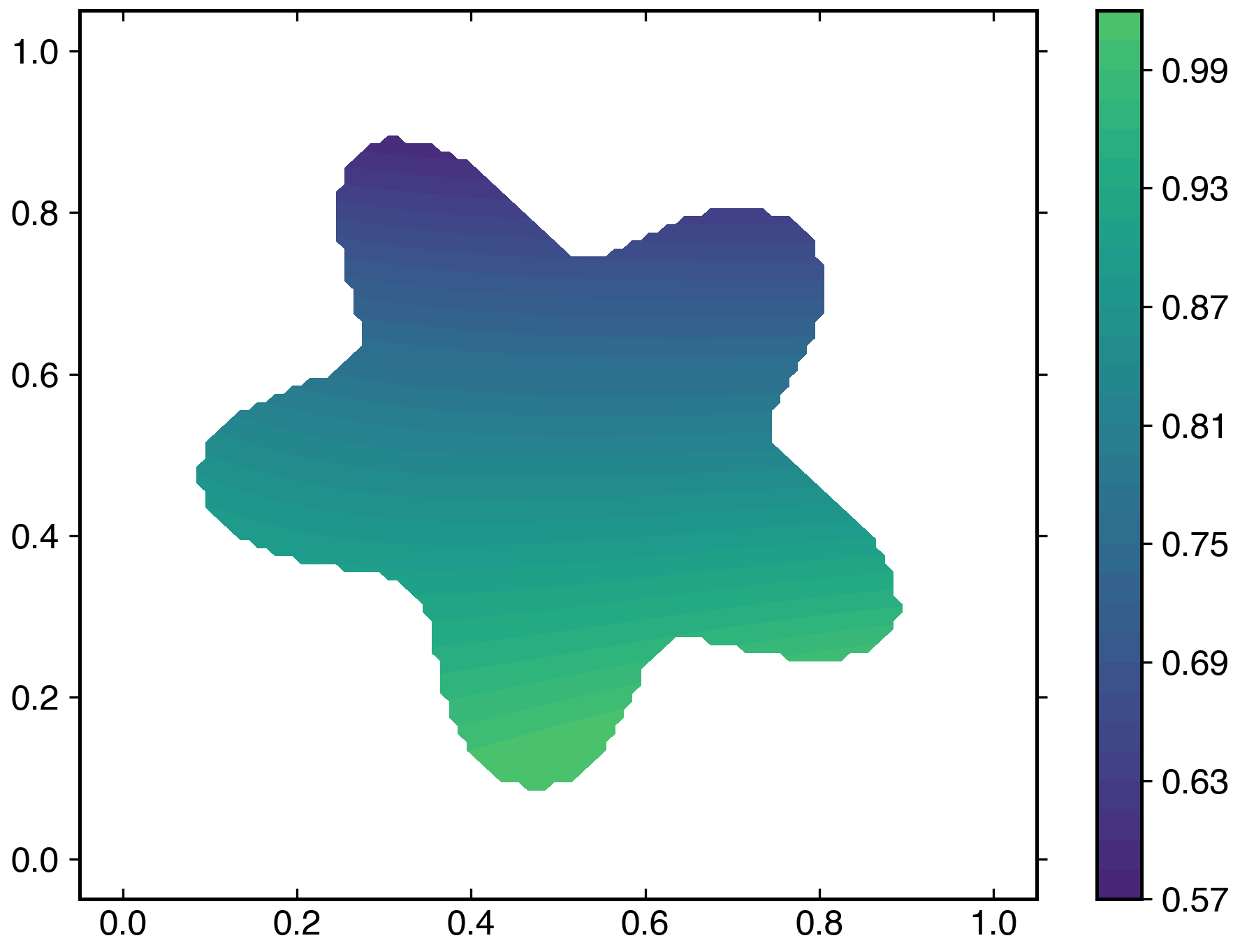}
      \caption{Exact distribution}
      \label{fig:var_coeff_gf_err_star}
  \end{subfigure}
\caption{The contour of the learned field using BIN-G and the error distribution for the Helmholtz equation with variable variable $\sigma$ and $\kappa$.}
     \label{fig:varcoeff_force_sols}
\end{figure}

In \cref{fig:var_coeff_gf}, we plot the learned Green's function. We plot the $L_2$ error
in the test points using the learned Green's function for different domain
shapes as done in previous test cases. Since the $\kappa$ value is variable
it involves lower eigenvalues as well. Therefore, the errors are higher
compared to other variable coefficient test cases. In the case of the
rectangular domain, the errors are higher due to corners compared to other
domain shapes. In \cref{fig:varcoeff_force_sols}, we plot the learned
solution with the exact distribution. Similar to other test case, the
rectangular domain shows a different contour line and magnitude. However,
in case of smooth domains the contours are close to the exact distribution.

\section{Conclusions}
\label{sec:conclusions}

In this paper, we propose a novel boundary integral network with unknown
Green's function referred to as BIN-G. In the BIN-G, we employ a radial basis
function (RBF) kernel-based neural network to evaluate the Green's
function, and multi-layer perceptron networks to evaluate the density
functions. The parameters of the RBF-based Green's function network viz.\
particle position and smoothing length are closely related to the sampling
space. Therefore, a careful initialization of the particle position and
smoothing length offer faster learning near the singularity of the Green's
function. We use the spherical symmetry feature of the domain-independent
Green's function in our network enabling us to learn Green's function
using a one-dimensional sample space.

We train the neural network by simultaneously minimizing the residual of
the PDE and the mean-squared error of the solution using the boundary
integral equations. We use a much larger sample space on which PDE residual
is minimized, compared to the domain on which solution to prescribed test
functions is learned using the BIN-G. Therefore, the learned Green's
function is defined for a very large space and is not constrained by the
homogeneous boundary condition. Hence, it can be readily employed for any
domain shape, boundary condition, and forcing function.

The efficacy of the proposed method to solve PDEs, for which the ground
truth Green's functions are known, has been demonstrated. We use the
Laplace and Helmholtz equations to validate our method. We apply the method
to learn domain-independent Green's function of variable coefficient PDEs,
for which analytical forms of Green's function are not available. We
demonstrate the applicability of the proposed network by solving a
different problem on different domain shapes using the learned Green's
function.

It is possible to enhance the obtained accuracy through the implementation
of an adaptively sampled domain, strategically placing refined samples in
proximity to the boundary. This approach not only refines training around
singularities but also mitigates quadrature errors. However, these
modifications give rise to convergence challenges, which can be addressed in
the future. The proposed method can be extended to solve interface boundary
problems as well. In the future, we would like to extend the proposed
method to solve moving interface boundary problems.

\section{Acknowledgements}
\label{sec:ack}

S. L and M. K acknowledge the support from the National Science
Foundation, ECCS-1927432. S. L also acknowledges the partial support from
the National Science Foundation,  grants DMS-1720420 and DMS-2309798. M. C  thanks NSF-1936873 and 2027725 for partial supports.

\bibliographystyle{model6-num-names}
\bibliography{references}

\end{document}